\documentclass[12pt]{article}
\usepackage{epsfig}
\usepackage{amssymb}
\usepackage{amsmath}
\usepackage{amsfonts}
\usepackage{graphicx}
\usepackage{mathrsfs}
\DeclareMathAlphabet{\mathscrbf}{OMS}{mdugm}{b}{n}
\usepackage{mathabx}
\usepackage[dvips]{color}
\usepackage{multirow}
\usepackage{calc}
\usepackage{accents}

\makeatletter
\newcommand{\shorteq}{\mathrel{\mkern0.2mu\mathpalette\shorteq@\relax\mkern0.2mu}}
\newcommand{\shorteq@}[2]{\scalebox{0.5}[1]{$\m@th#1=$}}

\newcommand{\longeq}[1]{\mathrel{\mathpalette\longeq@{#1}}}
\newcommand{\longeq@}[2]{%
  \begingroup
  \sbox\z@{$\m@th#1=$}%
  \ifdim#2<\wd\z@
    \resizebox{#2}{\height}{\box\z@}%
  \else
    \ifdim#2<3\wd\z@
      \hbox to #2{$\m@th#1=\hss=\hss=\hss=$}%
    \else
      \hbox to #2{$\m@th#1=\cleaders\hbox to 0.2\wd\z@{\hss$#1=$\hss}\hfil=$}%
    \fi
  \fi
  \endgroup
}
\makeatother

\newcommand{\bsigma}{\boldsymbol{\sigma}}

\newcommand{\R}{\mathbb{R}}
\newcommand{\C}{\mathbb{C}}

\newcommand{\fc}{\mathfrak{c}}

\newcommand{\ff}{\mathfrak{f}}
\newcommand{\fg}{\mathfrak{g}}

\newcommand{\fn}{{\mathfrak{n}}}

\newcommand{\fs}{\mathfrak{s}}

\newcommand{\fz}{\mathfrak{z}}

\newcommand{\fK}{\mathfrak{K}}

\newcommand{\bk}{\mathbf{k}}

\newcommand{\bfr}{\mathbf{r}}

\newcommand{\bv}{\mathbf{v}}

\newcommand{\bF}{\mathbf{F}}
\newcommand{\bG}{\mathbf{G}}

\newcommand{\bI}{\mathbf{I}}

\newcommand{\bM}{\mathbf{M}}

\newcommand{\cK}{\mathcal{K}}

\newcommand{\cO}{\mathcal{O}}
\newcommand{\cP}{\mathcal{P}}

\newcommand{\cR}{\mathcal{R}}

\newcommand{\cT}{\mathcal{T}}

\newcommand{\cX}{\mathcal{X}}
\newcommand{\cY}{\mathcal{Y}}

\newcommand{\be}{\begin{equation}}
\newcommand{\ee}{\end{equation}}
\newcommand{\bea}{\begin{eqnarray}}
\newcommand{\eea}{\end{eqnarray}}
\newcommand{\nn}{\nonumber}
\newcommand{\kt}{\rangle}
\newcommand{\br}{\langle}

\newcommand{\ed}{\end{document}}

\newcommand{\bi}{\begin{itemize}}
\newcommand{\ei}{\end{itemize}}

\newcommand{\bce}{\begin{center}}
\newcommand{\ece}{\end{center}}

\newcommand{\sD}{\mathscr{D}}

\newcommand{\sF}{\mathscr{F}}
\newcommand{\sG}{\mathscr{G}}

\newcommand{\bsH}{\mathscrbf{H}}

\newcommand{\sT}{\mathscr{T}}
\newcommand{\sV}{\mathscr{V}}
\newcommand{\sW}{\mathscr{W}}

\newcommand{\bsU}{\mathscrbf{U}}

\newcommand{\bcK}{{\boldsymbol{\cK}}}

\newcommand{\bzero}{{\boldsymbol{0}}}

\newcommand{\for}{{\mbox{\rm for}}}

\newcommand{\sinc}{{\rm sinc}}

\begin{document}

\title{Dynamical formulation of low-frequency scattering\\ in two and three dimensions}



\author{Farhang Loran\thanks{E-mail address: loran@iut.ac.ir}~ and
Ali~Mostafazadeh\thanks{Corresponding author, e-mail address:
amostafazadeh@ku.edu.tr}\\[6pt]
$^{*}$Department of Physics, Isfahan University of Technology, \\ Isfahan 84156-83111, Iran\\[6pt]
$^\dagger$Departments of Mathematics and Physics, Ko\c{c}
University,\\  34450 Sar{\i}yer, Istanbul, T\"urkiye}

\date{ }
\maketitle

\begin{abstract}

The transfer matrix of scattering theory in one dimension can be expressed in terms of the time-evolution operator for an effective non-unitary quantum system. In particular, it admits a Dyson series expansion which turns out to facilitate the construction of the low-frequency series expansion of the scattering data. In two and three dimensions, there is a similar formulation of stationary scattering where the scattering properties of the scatterer are extracted from the evolution operator for a corresponding effective quantum system. We explore the utility of this approach to scattering theory in the study of the scattering of low-frequency time-harmonic scalar waves, $e^{-i\omega t}\psi(\mathbf{r})$, with $\psi(\mathbf{r})$ satisfying the Helmholtz equation, $[\nabla^2+k^2\hat\varepsilon(\mathbf{r};k)]\psi(\bfr)=0$, $\omega$ and $k$ being respectively the angular frequency and wavenumber of the incident wave, and $\hat\varepsilon(\mathbf{r};k)$ denoting the relative permittivity of the carrier medium which in general takes complex values. {We obtain explicit formulas for low-frequency scattering amplitude, examine their effectiveness in the study of a class of exactly solvable scattering problems, and outline their application in devising a low-frequency cloaking scheme.}


\end{abstract}

\section{Introduction}
\label{S1}

Scattering of low-frequency waves has been a focus of attention since the pioneering works of Lord Rayleigh on the scattering of optical and acoustic waves by small obstacles \cite{Strutt-1871,Strutt-1897}. Because of its wide range of applications \cite{Bernstein-1974,Bannister-1984,deBadereau-2003,Kaushal-2016,Cummer 2020,Omer-2020,Hartinger-2022}, the subject has been extensively studied by several generations of physicists  \cite{Stevenson-1953,newton-1986,Chadan-1998,Sjoberg-2009,Khuri-2009, Majic-2019}, mathematicians \cite{Kriegsmann-1983,Bolle-1985,Kleinman-1986, Klaus-1987,Kleinman-1994,Aktosun-2001}, and engineers \cite{Vazouras-2000,Chen-2002,Smith-2019,vanHelvoort-2021}. The basic results of these studies have been presented in monographs such as Ref.~\cite{newton-book,yafaev,dassios-book}. These mainly rely on the standard approach to stationary scattering which makes use of the Lippmann-Schwinger equation and Green's functions for the wave equation. In Refs.~\cite{jmp-2021,jpa-2021} we pursue a different approach whose central ingredients are the notion of transfer matrix \cite{sanchez,tjp-2020} and its recently discovered Dyson series expansion \cite{ap-2014,pra-2014a}. The latter turns out to provide an effective method of constructing the low-frequency expansion of the scattering data in one dimension. The purpose of the present article is to develop a similar approach to low-frequency scattering of scalar waves in two and three dimensions.

Consider time-harmonic waves, $e^{-i\omega t}\psi(\bfr)$, where $\psi:\R^d\to\C$ solves the Helmholtz equation, 
	\be
	\left[\nabla^2+k^2\hat\varepsilon(\bfr;k)\right]\psi(\bfr)=0, 
	\label{H-eq}
	\ee
$d$ is the dimension of the space, $\omega$ and $k$ are respectively the angular frequency and wavenumber of the incident wave, and $\hat\varepsilon:\R^d\times\R^+\to\C$ is the relative permittivity of the carrier medium, i.e., $\hat\varepsilon(\bfr;k)=\fn(\bfr;k)^2$, where $\fn:\R^d\times\R^+\to\C$ is the (possibly complex) refractive index of the medium. 

In two dimensions, (\ref{H-eq}) describes the propagation of time-harmonic transverse electric (TE) waves in a nonmagnetic material with translational symmetry along one of the transverse directions, i.e., when the relative permittivity of the medium is a function of two of the Cartesian coordinates, say $x$ and $y$, while the electric field is along the $z$ direction \cite{ol-2017}. Another area of application of (\ref{H-eq}) is acoustics. The wave equation for time-harmonic pressure waves propagating in a compressible fluid in two or three dimensions reduces to an equation of the form (\ref{H-eq}) when the fluid's density is constant \cite{colton-kress}.\footnote{{In this case $\hat\varepsilon:=|\fc_0/\fc|^2$, where $\fc$ is the speed of sound in the fluid and $\fc_0$ is its value at spatial infinity, \cite{colton-kress,Bergmann,Martin}. For this reason, in acoustics, $\hat\varepsilon$ is usually called ``square of the refractive index'' \cite{Martin}. Some authors call it ``refractive index'' \cite{colton-kress}.}} When both the density and compressibility (speed of wave) vary in space, the fluid's pressure satisfies Bergmann's equation which can be mapped to an equation of the form (\ref{H-eq}) by a change of variable \cite{Bergmann,Martin}.

\begin{figure}
        \begin{center}
        \includegraphics[scale=.45]{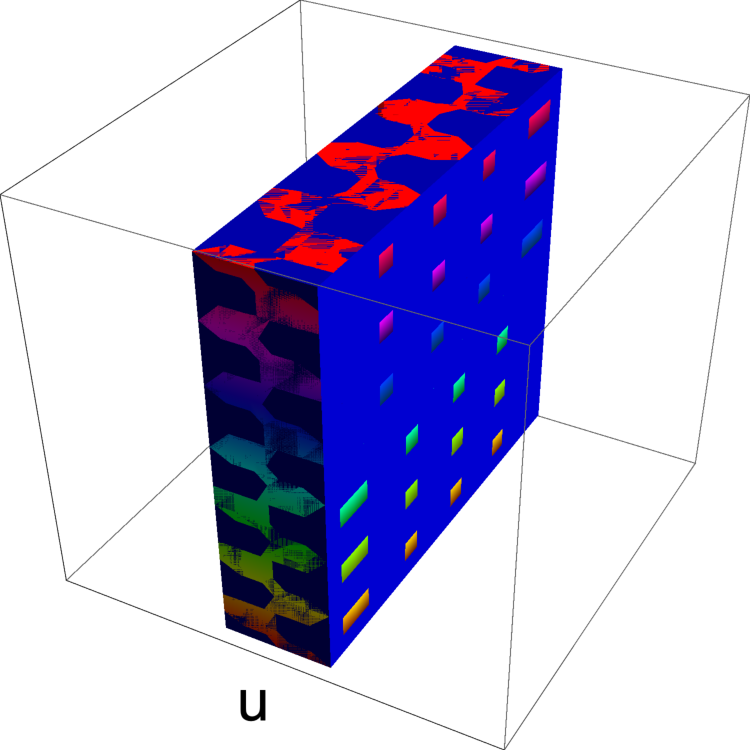} 
        \caption{Schematic views of an infinite planar slab containing an isotropic inhomogeneous material with possible regions of gain and loss.}
        \label{fig1}
        \end{center}
        \end{figure}

Suppose that $\hat\varepsilon(\bfr;k)=1$ outside an infinite planar slab of thickness $\ell$. Then we can choose a Cartesian coordinate system with one of the coordinate axes being normal to the slab's boundaries.
Let $u$ be the coordinate along this axis, as shown in Fig.~\ref{fig1}, 
and choose the origin of the coordinate system such that $\hat\varepsilon(\bfr;k)=1$ for $u\notin[0,\ell]$. We wish to study the scattering of incident plane waves of wavenumber $k$ much smaller than $\ell^{-1}$ with the source of the incident wave located on either of the planes $u=\pm\infty$, and the detectors measuring the amplitude of the scattered wave placed on both of these planes. More specifically, we are interested in constructing the low-frequency expansion of the scattering amplitude which is a series expansion in powers of $k\ell$. 

In Ref.~\cite{jpa-2021}, we show that the dynamical formulation of stationary scattering (DFSS) \cite{ap-2014,pra-2014a} provides an effective method of determining the low-frequency expansion of the reflection and transmission amplitudes for the above scattering problem in one dimension. This corresponds to situations where the slab containing the scattering medium has translational symmetry along both transverse directions to the $u$ axis. In Ref.~\cite{pra-2021}, we develop a DFSS for scalar waves propagating in two and three dimensions. In the present article, we examine its utility in the study of low-frequency scattering of scalar waves in these dimensions, i.e., when the slab has translational symmetry along one of the transverse directions to the $u$ axis, or has no translational symmetry at all. 

It is important to notice that standard methods of low-frequency scattering apply for situations where the inhomogeneity of the medium is confined to a compact region of the space \cite{dassios-book}, i.e., the interaction potential has a finite range. The scattering setup we are considering in the present article applies more generally, for we do not impose this restriction on the behavior of $\hat\varepsilon(\bfr;k)$ inside the slab which is an unbounded region of space. 

The organization of the article is as follows. In Sec.~\ref{S2}, we provide a brief discussion of the basic ingredients of DFSS in one and two dimensions. In Sec.~\ref{S3}, we use it to address the above low-frequency scattering problem in two dimensions. {In Sec.~\ref{S4}, we examine the utility of the results of Sec.~\ref{S3} in the study of a class of exactly solvable scattering problems.  In Sec.~\ref{S5}, we use these results to devise a low-frequency cloaking scheme.} In Sec.~\ref{S6}, we extend {these results} to three dimensions, and in Sec.~\ref{S7} we present our concluding remarks.

\section{Dynamical formulation of stationary scattering in 1D and 2D}
\label{S2}

\subsection{DFSS and low-frequency scattering in 1D}

Stationary scattering of scalar waves in one dimension admits a formulation in which the scattering data are extracted from a transfer matrix $\bM$. This is a complex $2\times 2$ matrix depending on the incident wavenumber $k$ whose entries $M_{ab}$ determine the left and right reflection and transmission amplitudes of the system, $R^{l/r}$ and $T^{l/r}$, according to \cite{Springer-book-2018},
	\begin{align}
	&R^l=-\frac{M_{21}}{M_{22}}, && R^r=\frac{M_{12}}{M_{22}},
	&&T^l=\frac{\det\bM}{M_{22}}, && T^r=\frac{1}{M_{22}}.
	\nn
	\end{align}
When the scatterer is modeled using an interaction potential or a permittivity profile, $\det\bM=1$, which establishes transmission reciprocity, $T^l=T^r$, \cite{tjp-2020,Springer-book-2018}. 

The transfer matrix $\bM$ has been known as a powerful tool for performing scattering calculations for multilayer and locally periodic scatterers since the 1940's \cite{jones-1941,abeles,thompson,yeh,pereyra,griffiths}. An intriguing property of $\bM$, which was noticed as late as 2014, is that it can be expressed in terms of the time-evolution operator for a non-unitary two-level quantum system \cite{ap-2014,pra-2014a}.  More precisely, there is a $2\times 2$ non-Hermitian matrix Hamiltonian $\bsH(\tau)$ with time-evolution operator, $\bsU(\tau,\tau_0):=\sT\left[-i\int_{\tau_0}^\tau d\tau\,\bsH(\tau)\right]$, such that $\bM=\bsU(+\infty,-\infty)$. Here $\tau$ is an effective evolution parameter which we can  identify with the spatial coordinate $x$, and $\sT$ is the corresponding time-ordering operation \cite{tjp-2020}. For the slab system of our interest in one dimension, with $\tau:=x$,
	\begin{align}
	&\bsH(x):=-\frac{k[\hat\varepsilon(x;k)-1]}{2}\left[\begin{array}{cc}
	1&e^{-2ikx}\\
	-e^{2ikx}&-1\end{array}\right].
	\nn
	\end{align}
Because $\hat\varepsilon(x;k)=1$ for $x\notin[0,\ell]$, $\bsH(x)$ vanishes for $x\notin[0,\ell]$, and we have
	\begin{align}
	\bM&=\sT\left[e^{-i\int_{0}^\ell dx\,\bsH(x)}\right]\nn\\
	&=\bI+\sum_{n=1}^\infty(-i)^n
	\int_0^\ell dx_n\int_0^{x_n}dx_{n-1}\cdots\int_0^{x_2}dx_1
	\bsH(x_n)\bsH(x_{n-1})\cdots\bsH(x_1).
	\label{dyson-1D}
	\end{align}

Now, suppose that $\hat\varepsilon(x;k)=1+w(\frac{x}{\ell};k)$ for a bounded function $w:\R\times\R^+\to\C$, and introduce
	\begin{align}
	\check{\bsH}(\check x):=
	-\frac{k\,w(\check x;k)}{2}
	\left[\begin{array}{cc}
	1&e^{-2ik\ell\check x}\\
	-e^{2ik\ell\check x}&-1\end{array}\right],\quad\quad\quad \check x:=\frac{x}{\ell}.
	\label{H-prime-1D}
	\end{align}
Then we can express (\ref{dyson-1D}) in the form
	\begin{align}
	\bM&=\bI+\sum_{n=1}^\infty(-ik\ell)^n
	\int_0^1 d\check x_n\int_0^{\check x_n}d\check x_{n-1}\cdots\int_0^{\check x_2}d\check x_1
	\check{\bsH}(\check x_n)\check{\bsH}(\check x_{n-1})\cdots\check{\bsH}(\check x_1).
	\label{dyson-1Dn}
	\end{align}
According to (\ref{H-prime-1D}), the $N$-th term on the right-hand side of (\ref{dyson-1Dn}) is given by $(k\ell)^N$ times a matrix-valued analytic function of $k\ell$. This provides the basic motivation for using (\ref{dyson-1Dn}) to construct the low-frequency series expansion of the transfer matrix  \cite{jpa-2021}.

\subsection{DFSS and the fundamental transfer matrix in 2D}

Let us adopt a Cartesian coordinate system  in which $u=x$, and suppose that the relative permittivity of our slab does not depend on $z$, i.e., the slab has translational symmetry along the $z$ axis. Then the scattering problem we wish to study is effectively two-dimensional. In Ref.~\cite{pra-2021}, we outline a DFSS in two dimensions where the scattering data are extracted from an analog of the transfer matrix $\bM$. This object, which we call the fundamental transfer matrix, is a $2\times 2$ matrix $\widehat\bM$ whose entries $\widehat M_{ab}$ are certain linear (integral) operators acting in an infinite-dimensional function space, i.e., $\widehat\bM$ is not a numerical matrix.  An important property of the fundamental transfer matrix is that similarly to its one-dimensional predecessor, it admits a Dyson series expansion. In the following we describe $\widehat\bM$ and recall some of its basic properties.

First, we introduce some notation. We label the null (zero) and identity matrices of all sizes by $\bzero$ and $\bI$, respectively. Given positive integers $m$ and $n$, we use $\C^{m\times n}$ and $\sF^m$ to respectively denote the vector spaces of $m\times n$ complex matrices and $m$-component complex-valued functions (tempered distributions) $\bF:\R\to\C^{m\times 1}$, so that for all $p\in\R$,
	\[\bF(p)=\left[\begin{array}{c}
	F_1(p)\\
	F_2(p)\\
	\vdots\\
	F_m(p)\end{array}\right].\]
Let $\sF^m_k$ be the subspace of $\sF^m$ consisting of functions whose {supports lie} in the interval $(-k,k)$, i.e., 
	\[\sF^m_k:=\left\{\bF\in\sF^m~|~\bF(p)=\bzero~\for~|p|\geq k\right\},\]
$\widehat I$ denote the identity operator acting in $\sF^m$, and $\widehat y,\widehat p,\widehat\varpi,\widehat\Pi_k:\sF^m\to\sF^m$ be the linear operators given by
	\begin{align}
	&(\widehat y\: \bF)(p):=i\partial_p\bF(p),
	\quad\quad\quad\quad (\widehat p\:\bF)(p)=p\,\bF(p),
	\quad\quad\quad\quad \widehat\varpi:=\varpi(\widehat p),
	\nn\\[6pt]
	&\varpi(p):=\left\{\begin{array}{ccc}
	\sqrt{k^2-p^2}&\for&|p|<k,\\
	i\sqrt{p^2-k^2}&\for&|p|\geq k,\end{array}\right.
	\quad\quad
	(\widehat\Pi_k \bF)(p):=\left\{\begin{array}{ccc}
	\bF(p)&\for&|p|<k,\\
	\bzero&\for&|p|\geq k.\end{array}\right.\nn
	\end{align}
Note that $\widehat\Pi_k$ is the projection operator mapping $\sF^m$ onto $\sF^m_k$. Using Dirac's bra-ket notation, we can express it in the form,
	\be
	\widehat\Pi_k=\int_{-k}^kdp\:|p\kt\,\br p|.
	\label{project}
	\ee

For the system we consider, the fundamental transfer matrix is the linear operator $\widehat\bM:\sF^2\to\sF^2$ given by 
	\be
	\widehat\bM=\widehat\Pi_k \sT\left[e^{-i\int_0^\ell dx~\widehat\bsH(x)}\right]\widehat\Pi_k,
	\label{M=}
	\ee
where 
	\begin{align}
	&\widehat\bsH(x):=\frac{1}{2}e^{-ix\widehat\varpi\bsigma_3}\widehat \sV(x)\widehat\varpi^{-1}\bcK\, e^{ix\widehat\varpi\bsigma_3},
	\label{H=2D}\\[6pt]
	&\widehat\sV(x):=v(x,\widehat y;k),\quad\quad
	v(x, y;k):=k^2[1-\hat\varepsilon(x,y;k)],
	\label{v-V=}\\[6pt]
	&\bsigma_3:=\left[\begin{array}{cc}
	1 & 0\\
	0 & -1\end{array}\right],\quad\quad	
	\bcK:=\left[\begin{array}{cc}
	1 & 1\\
	-1 & -1\end{array}\right].
	\label{K=}
	\end{align}
If we denote the Fourier transform of $v(x,y;k)$ with respect to $y$ by $\tilde v(x,p;k)$, i.e., set
	\[\tilde v(x,p;k):=\int_{-\infty}^\infty dy\:e^{-ipy}v(x,y;k),\] 
we can identify $\widehat\sV(x)$ with the integral operator,
	\be
	\left(\widehat\sV(x)\bF\right)(p)=\frac{1}{2\pi}\int_{-\infty}^\infty dq~\tilde v(x,p-q;k)\bF(q).
	\label{V=}
	\ee
This shows that $\widehat\bsH(x)$ and $\widehat\bM$ are integral operators acting in $\sF^2$. Note that because $\widehat\Pi_k$ projects $\sF^2$ onto $\sF^2_k$, (\ref{M=}) allows us to view $\widehat\bM$ as an integral operator mapping $\sF_k^2$ to $\sF_k^2$.
	
Next, we describe the relevance of $\widehat\bM$ to our scattering problem. To this end, we first recall the definition of the scattering amplitude.

\begin{figure}
        \begin{center}
        \includegraphics[scale=.25]{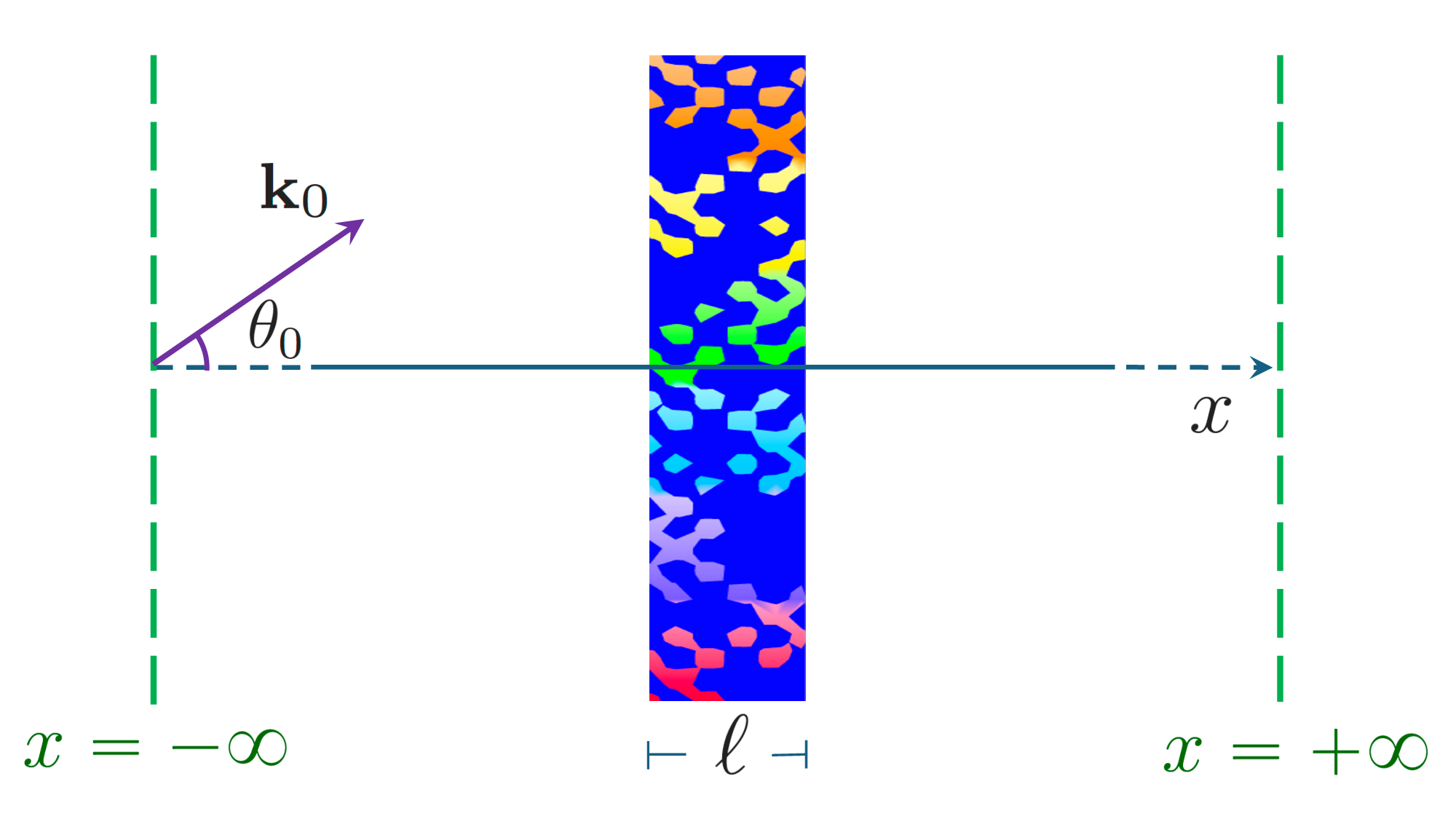}~~~~ 
        \includegraphics[scale=.25]{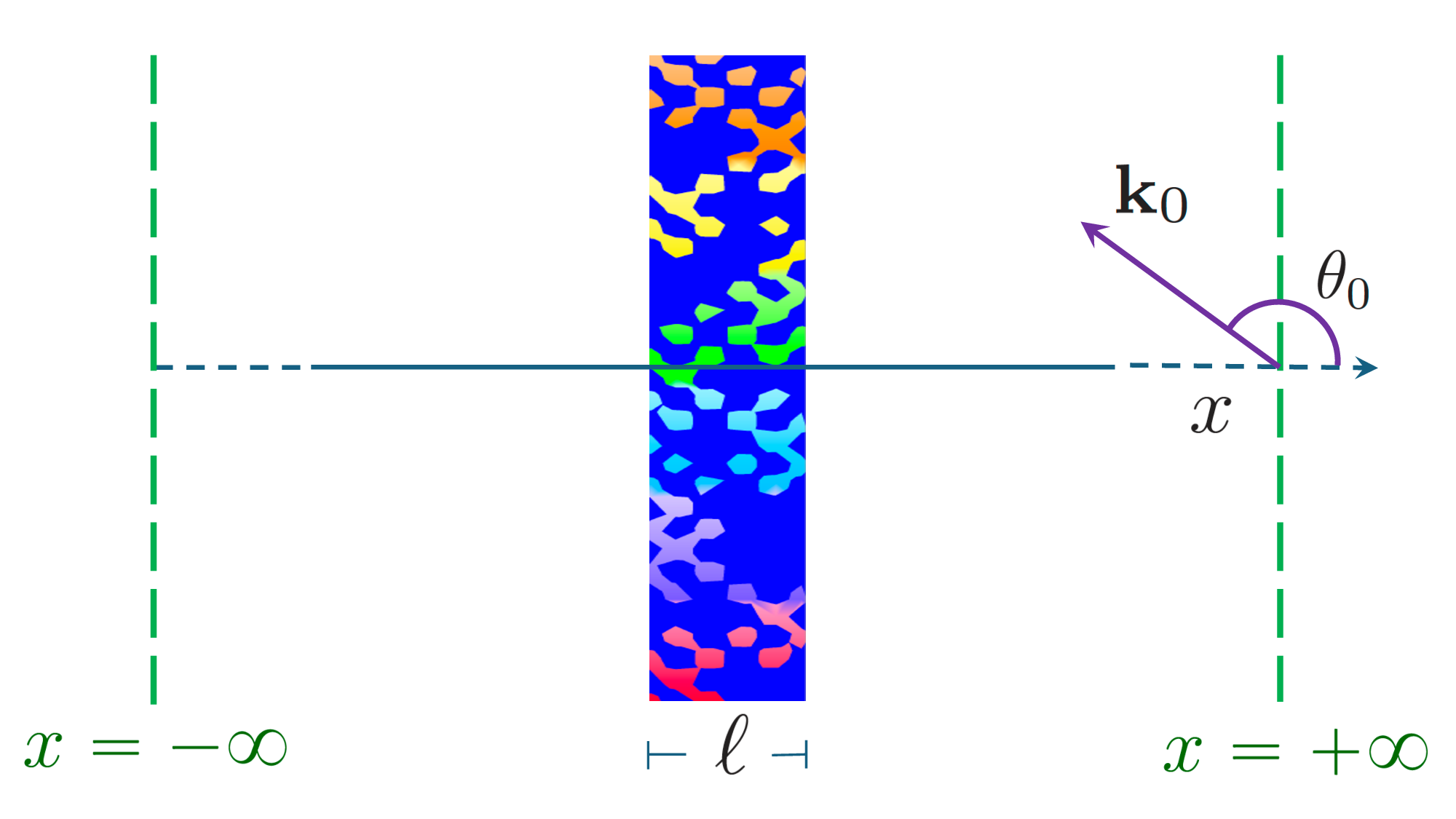} 
        \caption{Schematic views of the scattering setup for the scattering of left- and right-incident waves (respectively on the left and right) by an infinite planar slab containing an isotropic inhomogeneous material with possible regions of gain and loss. The green dashed lines correspond to $x=\pm\infty$ where the detectors are located.}
        \label{fig2}
        \end{center}
        \end{figure}

Since the detectors reside on the lines $x=\pm\infty$ and $\hat\varepsilon(x,y;k)=1$ for $x\notin[0,\ell]$, we are interested in the asymptotic behavior of the scattering solutions {\cite{yafaev}} of the Helmholtz equation (\ref{H-eq}) at $x=\pm\infty$. This is given by
	\be
	\psi(x,y)\to\frac{1}{2\pi}\left[e^{i\bk_0\cdot\bfr}+\sqrt{\frac{i}{kr}}\,e^{ikr}\ff(\theta)\right]
	~~\for~~x\to\pm\infty,
	\label{eq-11}
	\ee
where $r$ and $\theta$ are the polar coordinates of the position $\bfr$ of a generic detector, $\bk_0:=k(\cos\theta_0,\sin\theta_0)$ is the incident wave vector, $\theta_0$ is the incidence angle, and $\ff(\theta)$ is the scattering amplitude \cite{adhikari}.{\footnote{{Because in our scattering setup the detectors reside on the lines $x=\pm\infty$ as shown in Fig.~\ref{fig2}, the polar angle $\theta$ in (\ref{eq-11}) can take any value other than $\pm \frac{\pi}{2}$. For this reason, $x\to\pm\infty$ if and only if $r\to\infty$, and our definition of the scattering amplitude agrees with it's standard definition \cite{adhikari}.}}}

Recalling that the source of the incident wave is placed on either of the lines $x=\pm\infty$, we have $\pm\cos\theta_0>0$. We refer to incident waves with $\cos\theta_0>0$ and $\cos\theta_0<0$ as the left- and right-incident waves, and denote the corresponding scattering amplitudes respectively by $\ff^l(\theta)$ and $\ff^r(\theta)$. This means that 
	\be
	\ff(\theta)=\left\{\begin{array}{ccc}
	\ff^l(\theta)&\for&\cos\theta_0>0,\\
	\ff^r(\theta)&\for&\cos\theta_0<0.\end{array}\right.
	\label{f=ff}
	\ee
Fig.~\ref{fig2} provides a schematic description of the scattering of left- and right-incident waves. 
 
In Ref.~\cite{pra-2021}, we derive the following formulas for $\ff^{l/r}(\theta)$.
	\begin{align}
	\ff^l(\theta)=\frac{-i}{\sqrt{2\pi}}\times\left\{
	\begin{array}{ccc}
	A^l_+(k\sin\theta)-2\pi\delta(\theta-\theta_0)&\for&
	\cos\theta>0,\\
	B^l_-(k\sin\theta)&\for&
	\cos\theta<0,\end{array}\right.
	\label{f-Left}\\[6pt]
	\ff^r(\theta)=\frac{-i}{\sqrt{2\pi}}\times\left\{
	\begin{array}{ccc}
	A^r_+(k\sin\theta)&\for&
	\cos\theta>0,\\
	B^r_-(k\sin\theta)-2\pi\delta(\theta-\theta_0)&\for&
	\cos\theta<0,
	\end{array}\right.
	\label{f-Right}
	\end{align}
where $A^{l/r}_+$ and $B^{l/r}_-$ are the coefficient functions belonging to $\sF_k^1$ that satisfy
	\begin{align}
	&A^l_+=\widehat M_{12}B^l_-+\widehat M_{11}\check\delta_{p_0},
	\label{A-L}\\
	& \widehat M_{22} B^l_-=-\widehat M_{21}\check\delta_{p_0},
	\label{B-L}\\
	&A^r_+=\widehat M_{12}B^r_-,
	\label{A-R}\\
	&\widehat M_{22} B^r_-=\check\delta_{p_0},
	\label{B-R}
	\end{align}
and 
	\begin{align}
	&p_0:=k\sin\theta_0,
	&&\check\delta_{p_0}(p):=2\pi\varpi(p_0)\:\delta(p-p_0).
	\label{p0-selta=}
	\end{align}
Note that the conditions $\cos\theta<0$ and $\cos\theta>0$ in \eqref{f-Left} and \eqref{f-Right} correspond to the scattered waves reaching the detectors that are {respectively located at $x=-\infty$ and $x=+\infty$.}
	
Eqs.~(\ref{A-L}) and (\ref{A-R}) express $A^{l/r}_+$ in terms of $B^{l/r}_-$ whereas Eqs.~(\ref{B-L}) and (\ref{B-R}) are integral equations for $B^{l/r}_-$. We can solve the scattering problem provided that we determine the transfer matrix $\widehat\bM$ of the system and find the solution of these equations. As we see from (\ref{B-L}) and (\ref{B-R}),  the latter is equivalent to finding the inverse of 
$\widehat M_{22}$.\footnote{The situations where $\widehat M_{22}$ has a nontrivial kernel for certain real and positive values of $k$ is analogous to cases where the $M_{22}$ entry of the standard transfer matrix $\bM$ in 1D vanishes for some $k\in\R^+$. This marks the emergence of a spectral singularity \cite{prl-2009} which corresponds to the lasing threshold in gain media  \cite{pra-2011a}. See also \cite{longhi-2010,prsa-2012,pra-2013b,pra-2013d}.} 

It is easy to construct formal series solutions for (\ref{B-L}) and (\ref{B-R}). To do this, first we note that $B^{l/r}_-$ and $A^{l/r}_+$ belong to $\sF^1_k$, and $\widehat M_{ab}$ map $\sF_k^1$ to $\sF_k^1$. This shows that as far as Eqs. (\ref{A-L}) -- (\ref{B-R}) are concerned, we can view $\widehat M_{ab}$ as linear operators acting in $\sF^1_k$. Next, we let $\widehat N_{ab}:\sF^1_k\to\sF^1_k$ be the linear operators given by 
	\be
	\widehat N_{ab}:=\delta_{ab}\widehat I-\widehat M_{ab},
	\label{N-ab=}
	\ee
where $\widehat I$ stands for the identity operator acting in $\sF^1_k$. Then $\widehat M_{22}^{-1}=(\widehat I-\widehat N_{22})^{-1}=\sum_{j=0}^\infty\widehat N_{22}^j$, and we find the following series solutions of (\ref{B-L}) and (\ref{B-R}).		
	\begin{align}
	&B^l_-=\sum_{j=0}^\infty\widehat N_{22}^j \widehat N_{21}\check\delta_{p_0},
	&&B^r_-=
	\sum_{j=0}^\infty\widehat N_{22}^j \check\delta_{p_0}.
	\label{B-series}
	\end{align}
Substituting these in (\ref{A-L}) and (\ref{A-R}), we obtain
	\begin{align}
	&A^l_+=\check\delta_{p_0}-\widehat N_{11}\check\delta_{p_0}-
	\sum_{j=0}^\infty
	\widehat N_{12} \widehat N_{22}^j\widehat N_{21} \check\delta_{p_0},
	&&A^r_+=-\sum_{j=0}^\infty\widehat N_{12}\widehat N_{22}^j \check\delta_{p_0}.
	\label{A-series}
	\end{align}
Using Dirac's bra-ket notation, where $\br p|f\kt$ stands for $f(p)$, we can express (\ref{B-series}) and (\ref{A-series}) in the form:
	\begin{align}
	&B^l_-(p)=2\pi \varpi(p_0)
	\sum_{j=0}^\infty\br p|\widehat N_{22}^j \widehat N_{21}|p_0\kt,
	\label{B-L-series} \\
	&B^r_-(p)=2\pi \varpi(p_0)\Big[\delta(p-p_0)+
	\sum_{j=1}^\infty\br p|\widehat N_{22}^j |p_0\kt\Big],	
	\label{B-R-series} \\
	&A^l_+(p)=2\pi \varpi(p_0)\Big[\delta(p-p_0)-\br p|\widehat N_{11}|p_0\kt-
	\sum_{j=0}^\infty
	\br p|\widehat N_{12} \widehat N_{22}^j\widehat N_{21}|p_0\kt\Big],
	\label{A-L-series} \\
	&A^r_+(p)=-2\pi \varpi(p_0)\sum_{j=0}^\infty\br p|\widehat N_{12}\widehat N_{22}^j |p_0\kt,
	\label{A-R-series} 
	\end{align}
where $p\in(-k,k)$. Note that because $\varpi(p_0)=k|\cos\theta_0|$, (\ref{p0-selta=}) implies
	\begin{align}
	\check\delta_{p_0}(k\sin\theta)=2\pi\delta(\theta-\theta_0)~~~\for~~~\cos\theta\cos\theta_0>0.
	\label{id-0}
	\end{align}

\subsection{An alternative representation of the Dyson series for $\widehat\bM$}

According to (\ref{M=}), the determination of the fundamental transfer matrix is equivalent to the evaluation of the time-ordered exponential,
	\be
	 \sT\left[e^{-i\int_{x_0}^x dx~\widehat\bsH(x)}\right]:=
	 \widehat I+\sum_{n=1}^\infty (-i)^n
	\int_{x_0}^x dx_n\int_{x_0}^{x_n}dx_{n-1}\cdots\int_{x_0}^{x_2}dx_1
	\widehat\bsH(x_n)\widehat\bsH(x_{n-1})\cdots\widehat\bsH(x_1).
	\label{dyson-2D}
	\ee
In the following, we offer an expression for the right-hand side of this relation which makes its dependence on the potential $v$ more explicit.

Let $\sG^m$ denote the set of functions $\bG:\R\to\sF^m$, so that for all $x,p\in\R$, $\bG(x)\in\sF^m$ and $\big(\bG(x)\big)(p)\in\C^{m\times 1}$, and $\Theta:\R\to\R$ and $\sinc:\R\to\R$ be respectively the Heaviside step function and the sinc function, which are defined by
	\begin{align}
	&\Theta(x):=\left\{\begin{array}{ccc}
	1 & \for & x\geq 0,\\
	0 & \for & x<0,\end{array}\right.
	&&\sinc(x):=\sum_{j=0}^\infty\frac{(-1)^j x^{2j}}{(2j+1)!}=
	\left\{\begin{array}{ccc}
	x^{-1}\sin x &\for&x\neq 0,\\
	1 &\for&x=0.\end{array}\right.
	\label{step}
	\end{align}
Furthermore, let $\widehat s(x):\sF^m\to\sF^m$ and $\widehat\sV(\widehat x),\widehat S:\sG^m\to\sG^m$ be the operators given by
	\begin{align}
	&\widehat s(x):=x\,\Theta(x)\,\sinc(x\widehat\varpi)
	\quad\quad\quad
	\big(\widehat\sV(\widehat x)\bG\big)(x)=\widehat\sV(x)\bG(x),
	\label{s-x=}\\
	&(\widehat S\bG)(x):=\int_{-\infty}^\infty dx'\; \widehat s(x-x')\bG(x')=
	\int_{-\infty}^x dx'\; \widehat s(x-x')\bG(x'),
	\end{align}	
where $\bG\in\sG^m$. Employing Dirac's bra-ket notation, we can express $\widehat\sV(\widehat x)$ and $\widehat S$ in the form,
	\begin{align}
	&\widehat\sV(\widehat x)=\int_{-\infty}^\infty dx\:|x\kt\, \widehat\sV(x)\,\br x|,
	\label{V-x=}\\
	&\widehat S=\int_{-\infty}^\infty dx\int_{-\infty}^\infty dx'\;|x\kt\,\widehat s(x-x')\,\br x'|
	=\int_{-\infty}^\infty dx\int_{-\infty}^x dx'\;|x\kt\,\widehat s(x-x')\,\br x'|.
	\label{S=}
	\end{align}
	
{In the appendix, we use \eqref{step} -- \eqref{S=} to compute the right-hand side of (\ref{dyson-2D}). The result is}
	\begin{align}
	\sT\left[e^{-i\int_{x_0}^x dx'\:\widehat\bsH(x')}\right]=\widehat I-\frac{i}{2}
	\int_{x_0}^x\!\! dx_n \int_{x_0}^{x_n}\!\!\! dx_1
	e^{-ix_n\widehat\varpi\bsigma_3}
	\widehat\cR(x_n,x_1)\,\widehat\varpi^{-1}\bcK\,
	e^{ix_1\widehat\varpi\bsigma_3},
	\label{compact}
	\end{align}
where
	\begin{align}
	\widehat\cR(x,x')&:=\delta(x-x')\widehat\sV(x)+\sum_{s=1}^\infty
	\int_{-\infty}^\infty\!\! dx'_{s}\int_{-\infty}^\infty\!\! dx'_{s-1}
	\cdots \int_{-\infty}^\infty\!\! dx'_1\,\widehat\sV(x)\widehat\sV(x,x'_{s},x'_{s-1}\cdots,x'_1,x')
	\nn\\
	&=\sum_{s=0}^\infty
	\br x|\widehat\sV(\widehat x)\big[\widehat S\widehat\sV(\widehat x)\big]^s| x'\kt
	\stackrel{\rm formally}{\longeq{32pt}}
	\br x|\widehat\sV(\widehat x)\big[\widehat I-\widehat S\widehat\sV(\widehat x)\big]^{-1}|x'\kt.
	\label{Q=}
	\end{align}
It is easy to read off the entries of the right-hand side of (\ref{compact}). This gives
	\be
	\sT\left[e^{-i\int_{x_0}^x dx'\:\widehat\bsH(x')}\right]_{ab}=
	\delta_{ab}\widehat I+
	\frac{i(-1)^a}{2}
	\int_{x_0}^x\!\! dx_2 \int_{x_0}^{x_2}\!\!\! dx_1
	e^{(-1)^a ix_2\widehat\varpi}
	\widehat\cR(x_2,x_1)\,\widehat\varpi^{-1}
	e^{(-1)^{b-1}ix_1\widehat\varpi},
	\label{entries}
	\ee
where $a,b\in\{1,2\}$, and we have changed the dummy integration variable $x_n$ in (\ref{compact}) to $x_2$.

In view of  (\ref{M=}) and (\ref{entries}), the entries of the fundamental transfer matrix have the form:
	\be
	\widehat M_{ab}=\delta_{ab}\widehat\Pi_k+
	\frac{i(-1)^a}{2}
	\int_{0}^\ell\!\! dx_2 \int_{0}^{x_2}\!\!\! dx_1\,\widehat\Pi_k
	e^{(-1)^a ix_2\widehat\varpi}\,\widehat\cR(x_2,x_1)\,\widehat\varpi^{-1}
	e^{(-1)^{b-1}ix_1\widehat\varpi}\widehat\Pi_k.
	\label{Mab=}
	\ee
Because the projection operator $\widehat\Pi_k$ commutes with functions of $\widehat p$, we have
	\be
	\widehat M_{ab}=\delta_{ab}\widehat\Pi_k+
	\frac{i(-1)^a}{2}
	\int_{0}^\ell\!\! dx_2 \int_{0}^{x_2}\!\!\! dx_1\,
	e^{(-1)^a ix_2\widehat\varpi}\widehat\Pi_k\,
	\widehat\cR(x_2,x_1)\widehat\Pi_k\,\widehat\varpi^{-1}
	e^{(-1)^{b-1}ix_1\widehat\varpi}.
	\label{Mab=2}
	\ee

\section{Low-frequency scattering in 2D}
\label{S3}

Suppose that there is a function $w:[0,\ell]\times\R\times\R^+\to\C$ such that
	\be
	\hat\varepsilon(x,y;k)=1+\left\{\begin{array}{ccc}
	w(\frac{x}{\ell},y;k) &\for& x\in[0,\ell],\\
	0&\for&x\notin[0,\ell],
	\end{array}\right.
	\label{epsilon=}
	\ee
and $w(\check x,y;k)$ is a bounded function of $\check x$.
Then, in view of (\ref{v-V=}), we have 
	\begin{align}
	&v(x,y;k)=\left\{\begin{array}{ccc}
	-k^2w(\frac{x}{\ell},y;k) &\for& x\in[0,\ell],\\
	0&\for&x\notin[0,\ell],
	\end{array}\right.
	&&\widehat\sV(x)=-k^2\widehat\sW(\mbox{\large$\frac{x}{\ell}$}),
	\label{cV=}
	\end{align}
where 
	\begin{align}
	&\widehat\sW(\check x):=\left\{\begin{array}{ccc}
	w(\check x,\widehat y;k)&\for& \check x\in[0,1],\\[3pt]
	\widehat 0&\for&\check x\notin[0,1].
	\end{array}\right.
	\label{cW=}
	\end{align}
	 	
Next, consider the linear operators $\widehat{\check s}(\check x):\sF^m\to\sF^m$ and $\widehat{\check S}:\sG^m\to\sG^m$ given by,
	\begin{align}
	\widehat{\check s}(\check x)&:=
	\ell^{-1}\widehat s(\ell\check x)=\check x\,\Theta(\check x)\,\sinc(\ell\check  x\widehat\varpi)
	=\sum_{j=0}^\infty (k\ell)^{2j}s_j(\check x)\left(1-k^{-2}\widehat p^2\right)^{j},
	\label{cs=}
	\\
	\widehat{\check S}&:=\int_{-\infty}^\infty d\check x\int_{-\infty}^\infty d\check x'\;|\check x\kt\,
	\widehat{\check s}(\check x-\check x')\br\check x'|,	
	\label{cS}
	\end{align}
where we have made use of (\ref{step}) and (\ref{s-x=}), and introduced the coefficient functions,
	\[s_j(\check x):=\frac{(-1)^j\check x^{2j+1}}{(2j+1)!}\,\Theta(\check x).\]
In view of (\ref{Q=}), (\ref{Mab=2}),  (\ref{cV=}),  (\ref{cs=}) and (\ref{cS}),
	\begin{align}
	&\widehat M_{ab}=\delta_{ab}\widehat\Pi_k+
	\frac{i(-1)^{a}}{2\ell}
	\int_{0}^1\!\! d\check x_2 \int_{0}^{\check x_2}\!\!\! d\check x_1\,
	e^{(-1)^a i\ell\check x_2\widehat\varpi}\widehat\Pi_k
	\widehat{{\check\cR}}(\check x_2,\check x_1)\widehat\Pi_k\,\widehat\varpi^{-1}
	e^{(-1)^{b-1}i\ell\check x_1\widehat\varpi},
	\label{Mab=3}
	\end{align}
where
	\begin{align}
	\widehat{{\check\cR}}(\check x,\check x')&:=
	\sum_{s=0}^\infty [-(k\ell)^2]^{s+1}\:\br \check x|
	\widehat{\sW}(\widehat x)\left[\widehat{\check S}\,\widehat{\sW}(\widehat x)\right]^s|\check x'\kt
	\nn \\
	&=-(k\ell)^2\delta(\check x-\check x')\widehat\sW(\check x)+
	\sum_{s=2}^\infty [-(k\ell)^2]^{s}\widehat\sW_s(\check x,\check x'),
	\label{Q=2}\\
	\widehat\sW_2(\check x,\check x')&:=		
	\widehat\sW(\check x)\,\widehat{\check s}(\check x-\check x')\,\widehat\sW(\check x'),
	\label{W-1}
	\end{align}
for $s\geq 3$ and $n\geq 3$,
	\begin{align}
	&\widehat\sW_s(\check x,\check x'):=	
	\int_{0}^{\check x}\!\! d\check x'_{s-2}\int_{0}^{\check x'_{s-2}}\!\! d\check x'_{s-3}
	\cdots \int_{0}^{\check x'_{2}}\!\! d\check x'_1\,
	\widehat\sW(\check x,\check x'_{s-2},\check x'_{s-3}\cdots,\check x'_1,\check x'),
	\label{W-s}\\
	&\widehat\sW(\check x_n,\check x_{n-1},\cdots,\check x_1):=
	\widehat\sW(\check x_{n})
	\widehat{\check s}(\Delta\check x_{n-1})\widehat\sW(\check x_{n-1})
	\widehat{\check s}(\Delta\check x_{n-2})
	\cdots\widehat\sW(\check x_{2})\widehat{\check s}(\Delta\check x_1)\widehat\sW(\check x_1),
	\label{W=c}
	\end{align}
and $\Delta\check x_j:=\check x_{j+1}-\check x_j$. It is important to notice that in addition to the explicit dependence of $\widehat{{\check\cR}}(\check x,\check x')$ on $k\ell$ through the terms $(k\ell)^2$ and $[-(k\ell)^2]^{s}$ on the right-hand side of (\ref{Q=2}), it also depends on $k\ell$ because $\widehat{\check s}(\check x)$, $\widehat\sW(\check x_n,\check x_{n-1}\,\cdots,\check x_1)$, and consequently $\widehat\sW_s(\check x,\check x')$ are functions of $k\ell$.  

Substituting (\ref{Mab=3}) in (\ref{N-ab=}), and noting that as an operator acting in $\sF^1_k$ the projection operator $\widehat\Pi_k$ coincides with the identity operator $\widehat 1$ for $\sF_k^1$, we obtain
	\begin{align}
	&\widehat N_{ab}=\frac{i(-1)^{a-1}}{2\ell}
	\int_{0}^1\!\! d\check x_2 \int_{0}^{\check x_2}\!\!\! d\check x_1\,
	e^{(-1)^a i\check x_2\ell\widehat\varpi}\widehat\Pi_k 
	\widehat{{\check\cR}}(\check x_2,\check x_1)\widehat\Pi_k\,\widehat\varpi^{-1}
	e^{(-1)^{b-1}i\check x_1\ell\widehat\varpi}.
	\label{Nab=3}
	\end{align}
To determine the low-frequency behavior of the scattering amplitudes $\ff^{l/r}(\theta)$, we explore the nature of the $k\ell$-dependence of the right-hand side of this equation. To this end, we make the following observations.
	\begin{enumerate}
	\item According to (\ref{project}),
		\begin{align}
		&e^{(-1)^a i\check x_2\ell\widehat\varpi}\widehat\Pi_k=
		\int_{-k}^kdp\:e^{(-1)^a i\check x_2 k\ell\sqrt{1-p^2/k^2}}\,|p\kt\,\br p|,
		\label{z11}\\
		&\widehat\Pi_k\,\widehat\varpi^{-1}e^{(-1)^{b-1}i\check x_1\ell\widehat\varpi}=
		\frac{1}{k}
		\int_{-k}^kdp'\: \frac{e^{(-1)^{b-1}i\check x_1 k\ell\sqrt{1-{p'}^{2}/k^2}}}{\sqrt{1-{p'}^{2}/k^2}}
	\,|p'\kt\,\br p'|,\\
		&\br p|\widehat\sW(\check x)|p'\kt=\frac{\tilde w(\check x,p-p';k)}{2\pi}~~~~\for~~~\check x\in[0,1],
		\end{align}
		where a tilde over a function stands for its Fourier transform with respect to $y$.
	
	\item Eqs.~(\ref{cs=}), (\ref{W-1}), and (\ref{W=c}), imply
		\begin{align}
		\widehat\sW_2(\check x,\check x')&=\sum_{j=0}^\infty (k\ell)^{2j}
		s_j(\check x-\check x')
		\widehat\sW(\check x)(1-k^{-2}\widehat p^2)^j\widehat\sW(\check x'),
		\label{W-2-series}\\
		\widehat\sW(\check x_n,\check x_{n-1},\cdots,\check x_1)=&
		\sum_{j_1,j_2,\cdots,j_{n-1}=0}^\infty 
		s_{j_{n-1}}(\Delta\check x_{j_{n-1}})
		s_{j_{n-2}}(\Delta\check x_{j_{n-2}})\cdots 
		s_{j_1}(\Delta\check x_1)
		\times\nn\\
		&\hspace{2.3cm}
		\widehat\sW_{j_1,j_2,\cdots,j_{n-1}}(\check x_n,\check x_{n-1},\cdots,\check x_1),
		\label{W-3-series}
		\end{align}
where $n\geq 3$, 
		\begin{align}
		\widehat\sW_{j_1,j_2,\cdots,j_{n-1}}(\check x_n,\check x_{n-1},\cdots,\check x_1):=&
		\left\{\prod_{r=1}^{n-1}(k\ell)^{2j_{n-r}}
		\widehat\sW(\check x_{n-r+1})
		(1-k^{-1}\widehat p^2)^{j_{n-r}}
		\right\}
		\,\widehat\sW(\check x_1),
		\label{W-j-n=}
		\end{align}
and $\prod_{r=1}^{n-1}$ stands for the ordered product whose terms are ordered from left to right in ascending values of $r$.
	\end{enumerate}
In view of (\ref{Q=2}) 
and (\ref{Nab=3}) -- (\ref{W-j-n=}), we can express $\widehat N_{ab}$ in the form
	\be
	\widehat N_{ab}=\sum_{j=1}^\infty (k\ell)^{j}
	\int_{-k}^k dp\int_{-k}^kdp'\: N^{(j)}_{ab}(p,p';k)|p\kt\,\br p'|,
	\label{N-ab=2}
	\ee
where $N^{(j)}_{ab}(p,p';k)$ are coefficients that do not depend on $k\ell$. 

According to (\ref{f=ff}) --  (\ref{f-Right}), (\ref{B-L-series}) --  (\ref{A-R-series}), and  (\ref{N-ab=2}), the scattering amplitude $\ff(\theta)$ admits a low-frequency series expansion in positive integer powers of $k\ell$. To derive explicit formulas for the coefficients of this expansion, we need to calculate $N^{(j)}_{ab}(p,p';k)$. This is easy for  $j\in\{1,2\}$, because only the first term on the right-hand side of (\ref{Q=2}) contributes. The result is
	\begin{align}
	N^{(1)}_{ab}(p,p';k)=&\frac{(-1)^a\, i~ \overline{\tilde w}_0(p-p';k)}{4\pi\sqrt{1-{p'}^2/k^2}},
	\label{N1=}\\
	N^{(2)}_{ab}(p,p';k)=&\frac{1}{4\pi}
	\left[(-1)^{a+b}-\sqrt{\frac{k^2-p^2}{k^2-{p' }^2}}\right]
	\:\overline{\tilde w}_1(p-p';k),
	\label{N2=}
	\end{align}
where
	\be
	\overline{\tilde w}_l(p;k):=\int_0^1d\check x\: \check x^l\:\tilde w(\check x,p;k)
	=\frac{1}{\ell^{l+1}}\int_{0}^{\ell}dx\,x^l[\tilde{\hat\varepsilon}(x,p;k)-1],
	\quad\quad\quad l\in\{0,1\}.
	\label{w-ell=}
	\ee
The calculation of $N^{(3)}_{ab}(p,p';k)$ is also not that difficult. It yields
	\begin{align}
	N^{(3)}_{ab}(p,p';k)=&\frac{i(-1)^{a-1}\,}{4\pi
	\sqrt{1-{p' }^2/k^2}}\Bigg\{	
	\overline{\tilde w}_2(p-p';k)
	\Big[1-\frac{p^2+{p' }^2}{2k^2}
	-(-1)^{a+b}\sqrt{\Big(1-\frac{p^2}{k^2}\Big)\Big(1-\frac{{p' }^2}{k^2}\Big)}\,\Big]
    \nn\\
    &+\int_{0}^1\!\! d\check x_2 \int_{0}^{\check x_2}\!\!\! d\check x_1
    (\check{x}_2-\check{x}_1)\tilde Q(x_1,x_2,p-p';k)
    \Bigg\},
    \label{N3=}
	\end{align}
where $\tilde Q(x_1,x_2,p;k)$ stands for the Fourier transform of the function defined by
	\[Q(x_1,x_2,y;k):=w(x_1,y;k)w(x_2,y;k)\]
with respect to $y$, i.e.,
	\[\tilde Q(x_1,x_2,p;k):=\int_{-\infty}^\infty\!\! dy\: e^{-iyp} w(x_1,y;k)w(x_2,y;k)=
	\frac{1}{2\pi}\int_{-\infty}^{\infty}\!\!dq\,
	\tilde w(\check x_1,p-q;k)\tilde w(\check x_2,q;k).\] 
Obtaining explicit formulas for $N^{(j)}_{ab}(p,p';k)$ with $j\geq 3$ requires expressing the right-hand side of (\ref{W-2-series}) and (\ref{W-j-n=}) in the form
	\[\sum_{J=0}^{J_n} 
	\widehat\sW^{\,(J)}_{j_1,j_2,\cdots,j_{n-1}}(\check x_n,\check x_{n-1},\cdots,\check x_1)\widehat p^{\,J},\]
where $J_n:=2\sum_{l=1}^{n-1}j_l$ and $\widehat\sW^{(J)}_{j_1,j_2,\cdots,j_{n-1}}(\check x_n,\check x_{n-1},\cdots,\check x_1)$ are operators that commute with $\widehat y$.
	
Substituting (\ref{N-ab=2}) in (\ref{B-L-series}) --  (\ref{A-R-series}) and making use of  (\ref{N1=}) and (\ref{N2=}) we can determine the leading-order and next-to-leading-order terms in the low-frequency series expansion of the coefficient functions $A^{l/r}(p)$ and $B^{l/r}(p)$. Inserting the resulting expressions in (\ref{f-Left}) and (\ref{f-Right}) and making use of (\ref{f=ff}), we find 
	\be
	\ff(\theta)=	\ff^{(1)}(\theta)\,k\ell+\ff^{(2)}(\theta)\,(k\ell)^2+\cO(k\ell)^3 ,
	\label{f=series-2D}
	\ee
where
{\begin{align}
	&\ff^{(1)}(\theta):=\frac{k}{2\sqrt{2\pi}}\:\overline{\tilde w}_0\big(k\,\fs(\theta,\theta_0);k\big),
	\label{ff1=}\\
	&\begin{aligned}
	\ff^{(2)}(\theta):=&\frac{ik}{2\sqrt{2\pi}}\Big[
	-\fc(\theta,\theta_0)\,\overline{\tilde w}_1\big(k\,\fs(\theta,\theta_0);k\big)+\nn\\
	&\hspace{1.5cm}\frac{k}{4\pi}\int_{-\frac{\pi}{2}}^{\frac{\pi}{2}} d\varphi \:
\overline{\tilde w}_0\big(k\,\fs(\theta,\varphi);k\big)\,\overline{\tilde w}_0\big(k\,\fs(\varphi,\theta_0);k\big)\Big],
	\end{aligned}
	\label{ff2=}\\
	& \fs(\theta,\theta_0):=\sin\theta-\sin\theta_0,
	\quad\quad\quad
	\fc(\theta,\theta_0):=\cos\theta-\cos\theta_0,
	\end{align}
and $\cO(k\ell)^j$ stands for terms of order $j$ and higher in powers of $k\ell$.{\footnote{{One can similarly use (\ref{B-L-series}) --  (\ref{A-R-series}) and (\ref{N-ab=2}) -- (\ref{N3=}) to {determine} terms of order $(k\ell)^3$ that contribute to $\ff(\theta)$. The resulting expression is too lengthy to be reported here.}}} We call the approximation that neglects $\cO(k\ell)^3$ the second-order low-frequency approximation. Similarly, by first-order low-frequency approximation we mean the approximation in which we neglect all but the linear term in $k\ell$ on the right-hand side of (\ref{f=series-2D}).}%


\section{{Application to a class of exactly solvable problems}}
\label{S4}

{In Refs.~\cite{pra-2021,pra-2019}, we use the dynamical formulation of stationary scattering in two and three dimensions to construct short-range complex potentials for which the first Born approximation gives the exact expression for the scattering amplitude when the incident wavenumber $k$ does not exceed a prescribed value $\alpha$. In two dimensions, these are potentials $v(x,y)$ satisfying $\tilde v(x,p)=0$ for $p\leq\alpha$. In Ref.~\cite{jpa-2024} we use the standard (Lippmann-Schwinger) approach to stationary scattering to show that for every $k$ there is some positive integer $N$ such that the scattering problem for this class of potentials is exactly solvable by the $N$-th order Born approximation. For $k\leq\alpha$, $N=1$, the first Born approximation is exact, and the scattering amplitude takes the form:
	\be
	\ff(\theta)=-\frac{\tilde{\tilde v}\big(k\,\fc(\theta,\theta_0),k\,\fs(\theta,\theta_0);k\big)}{2\sqrt{2\pi}}~~~\for~~~k\leq\alpha,
	\label{exact=}
	\ee
where $\tilde{\tilde v}(p_x,p_y;k)$ stands for the two-dimensional Fourier transform of $v(x,y;k)$, i.e.,
	\be
	\tilde{\tilde v}(p_x,p_y;k)=\int_{-\infty}^\infty dx\int_{-\infty}^\infty dy\:
	e^{-i(xp_x+yp_y)}v(x,y;k).
	\label{2ed-FT}
	\ee
In view of (\ref{cV=}), the same applies for the permittivity profiles of the form (\ref{epsilon=}) provided that
	\be
	\tilde w(\check x,p;k)=0~~~\for~~~p\leq \alpha.
	\label{Born-exact}
	\ee}%
	
{It is not difficult to show that whenever (\ref{Born-exact}) holds and $k\leq\alpha$, the integrand on the right-hand side of (\ref{ff2=}) vanishes identically, and we have
	\be
	\ff^{(2)}(\theta)=\frac{-ik\, \fc(\theta,\theta_0)\,\overline{\tilde w}_1\big(k\,\fs(\theta,\theta_0);k\big)}{2\sqrt{2\pi}} 
	~~~~\for~~~~k\leq\alpha.
	\label{ff2=special}
	\ee
According to (\ref{ff1=}) and \eqref{ff2=special}, $f^{(1)}(\theta)$ and $f^{(2)}(\theta)$ are respectively proportional to $\overline{\tilde w}_0$ and $\overline{\tilde w}_1$. Because these depend linearly on $w$, for $k\leq\alpha$, we can determine $f^{(1)}(\theta)$ and $f^{(2)}(\theta)$ by expanding the right-hand side of (\ref{exact=}) in powers of $k\ell$ and neglecting the cubic and higher order terms. {To do this, we first use} (\ref{cV=}), (\ref{w-ell=}), and \eqref{2ed-FT} to show that
	\begin{align}
	\tilde{\tilde v}(p_x,p_y;k)&=-k^2\int_0^\ell dx\: e^{-ix p_x}\tilde w(\mbox{$\frac{x}{\ell}$},p_y)
	\label{ttv=}\\
	&=-k^2\ell\left[\,\overline{\tilde w}_0(p_y;k)-i\ell p_x \overline{\tilde w}_1(p_y;k)\right]+
	\cO(k\ell)^3.\nn
	\end{align}
Substituting this relation in (\ref{exact=}), we recover (\ref{f=series-2D}) with $f^{(1)}(\theta)$ and $f^{(2)}(\theta)$ given by (\ref{ff1=}) and (\ref{ff2=}) respectively. This provides a nontrivial consistency check on our derivation of these equations.}

In the following, we demonstrate the accuracy of the first- and second-order low-frequency approximations for a specific permittivity profile of the form (\ref{epsilon=}) that fulfills (\ref{Born-exact}), namely the one corresponding to
	\be
	w(\check x,y;k)=\frac{\fz\,e^{i\alpha y}}{(y/L+i)^2},
	\label{ex1}
	\ee
where $\fz$ is a possibly complex constant, and $\alpha$ and $L$ are positive real parameters.

Substituting (\ref{ex1}) in (\ref{w-ell=}) and making use of the result in (\ref{ff1=}) and (\ref{ff2=}), we find
	\begin{align}
	&\overline{\tilde w}_0(p;k)=2\overline{\tilde w}_1(p;k)=
	 2\pi \,\fz\, L^2(\alpha-p) e^{L(\alpha-p)}\,\Theta\big(L(p-\alpha)\big),
	 \label{tw0=ex1}\\
	&\ff^{(1)}(\theta)=-\sqrt{\frac{\pi}{2}}\,\fz\,\fK^2\,
	\big[\fs(\theta,\theta_0)-\mbox{\large$\frac{\alpha}{k}$}\big]\,
	e^{\fK\,[\alpha/k-\fs(\theta,\theta_0)]}\,
	\Theta\big(\fs(\theta,\theta_0)-\mbox{\large$\frac{\alpha}{k}$}\big),
	 \label{f1=ex1}\\
	&\ff^{(2)}(\theta)=\frac{i}{2}\left[-\fc(\theta,\theta_0)\,\ff^{(1)}(\theta)+\sqrt{\frac{\pi}{2}}\,\fz^2\fK^4\,
	e^{\fK\,[2\alpha/k-\fs(\theta,\theta_0)]}\,\cX(\sin\theta,\sin\theta_0,\mbox{\large$\frac{\alpha}{k}$})\right],
	 \label{f2=ex1}
	\end{align}
where $\Theta(\cdot)$ is the Heaviside step function given in (\ref{step}), $\fK:=kL$, and 	
	\begin{align}
	\cX(\varsigma,\varsigma_0,\xi):=&\frac{1}{2}\,\Theta(\varsigma-\varsigma_0-2\xi)\,\Theta(\varsigma-\xi+1)\,\Theta(1-\xi-\varsigma_0)\,\times\nn\\
	&\Big\{[2(\xi-\varsigma)(\xi+\varsigma_0)-1]
	\big[\sin^{-1}(\varsigma-\xi)-
	\sin^{-1}(\varsigma_0+\xi)\big]+\nn\\
	&~~2(\varsigma+\varsigma_0)\Big[\sqrt{1-(\varsigma_0+\xi)^2}-
	\sqrt{1-(\varsigma-\xi)^2}\Big]+\nn\\
	&~~(\varsigma-\xi)\sqrt{1-(\varsigma-\xi)^2}-
	(\varsigma_0+\xi)\sqrt{1-(\varsigma_0+\xi)^2}\Big\},
	\end{align}
$\varsigma,\varsigma_0\in[-1,1]$ and $\xi\in\R^+$.	
For $\xi>1$, $\Theta(\varsigma-\varsigma_0-2\xi)=0$, which implies 
that $\cX(\varsigma,\varsigma_0,\xi)=0$ for $\xi\geq 1$. This shows that 
for $k\leq\alpha$, the second term on the right-hand side of (\ref{f2=ex1}) vanishes, i.e.,
	\be
	\ff^{(2)}(\theta)= -\frac{i}{2}\,\fc(\theta,\theta_0)\,\ff^{(1)}(\theta)
	~~~~\for~~~~k\leq\alpha.
	\label{f2=ex1-b} 
	\ee
	
{Next, we employ (\ref{exact=}) and (\ref{ttv=}) to determine the exact expression for the scattering amplitude for the system given by (\ref{ex1}) for $k\leq\alpha$. The result is 
	\begin{align}
	\ff(\theta)
	&=\left[\frac{i (e^{-ik\ell\,\fc(\theta,\theta_0)}-1)}{\fc(\theta,\theta_0)}\right] f^{(1)}(\theta)~~~\for~~~k\leq\alpha,
	\label{exact-ex1}
	\end{align}
where we have also benefitted from (\ref{tw0=ex1}) and (\ref{f1=ex1}).}

{For $k\leq\alpha/2$, $\fs(\theta,\theta_0)-\alpha/k\leq\sin\theta-\sin\theta_0-2\leq 0$ which in view of (\ref{f1=ex1}) and \eqref{exact-ex1} implies $\ff(\theta)=0$ for all $\theta$ and $\theta_0$. This is consistent with the general results on the invisibility of this class of permittivity profiles for $k\leq\alpha/2$, \cite{pra-2021,pra-2019,ptep-2024}.} {For $k\leq\alpha$, $\fs(\theta,\theta_0)-\alpha/k\leq\sin\theta-\sin\theta_0-1$. Consequently, $\Theta\big(\fs(\theta,\theta_0)-\mbox{\large$\frac{\alpha}{k}$}\big)=0$ if $0<\theta_0<\pi$ or $\pi<\theta<2\pi$, and according to  (\ref{f1=ex1}), (\ref{f2=ex1}), and \eqref{exact-ex1},
	\be
	\ff^{(1)}(\theta)=\ff^{(2)}(\theta)=\ff(\theta)=0~~~\for~~~0\leq\theta_0\leq\pi~~~{\rm or}~~~\pi\leq\theta\leq 2\pi.
	\nn
	\ee
This shows that in order to examine the validity of the first- and second-order low-frequency approximations, we should compare the behavior of $\ff^{(1)}(\theta), \ff^{(2)}(\theta)$, and $\ff(\theta)$ for $0<\theta<\pi$ and $\pi<\theta_0<2\pi$.}

\begin{figure}
	\begin{center}
        \includegraphics[scale=.35]{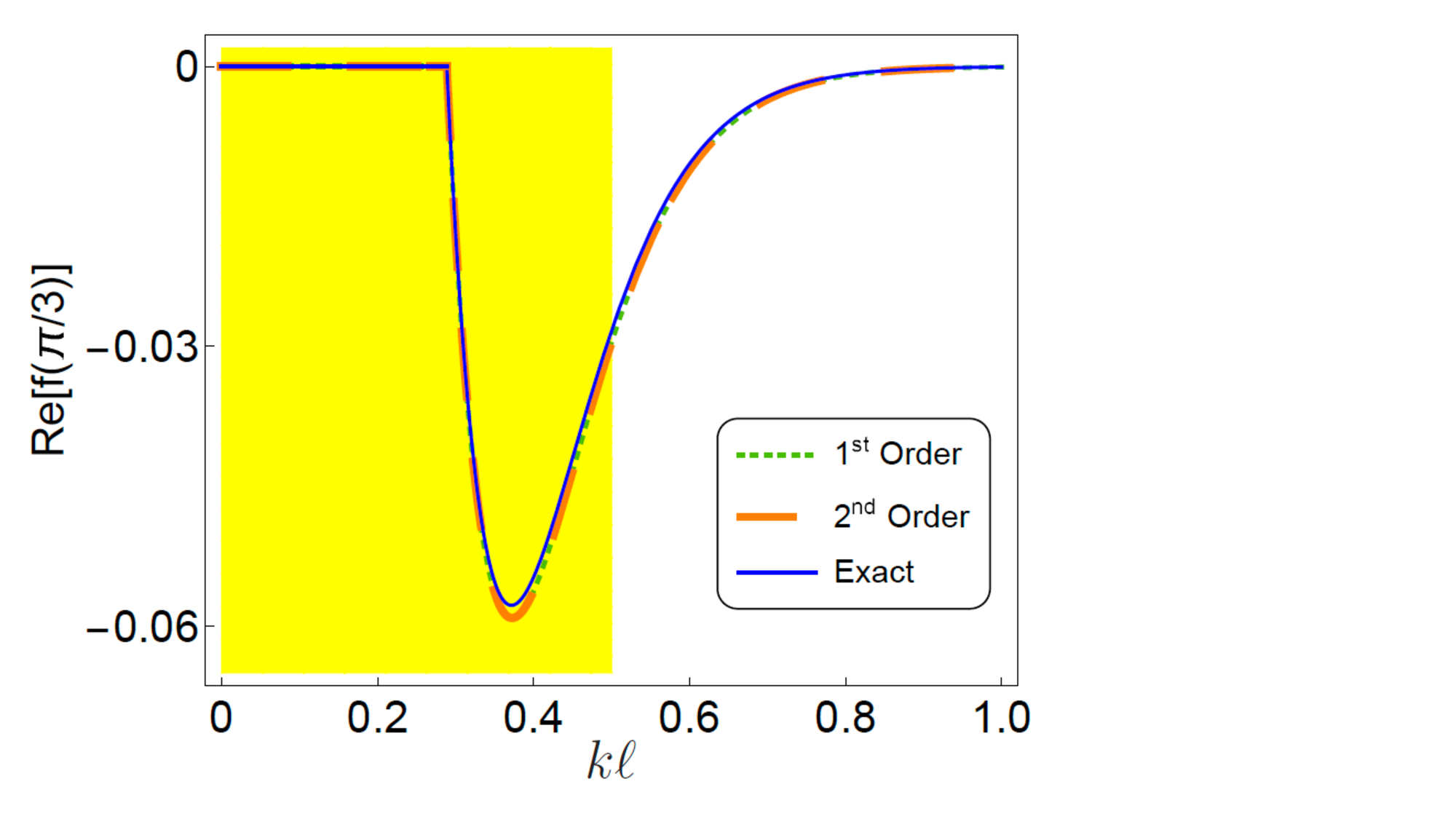}~\hspace{-3cm}~\includegraphics[scale=.35]{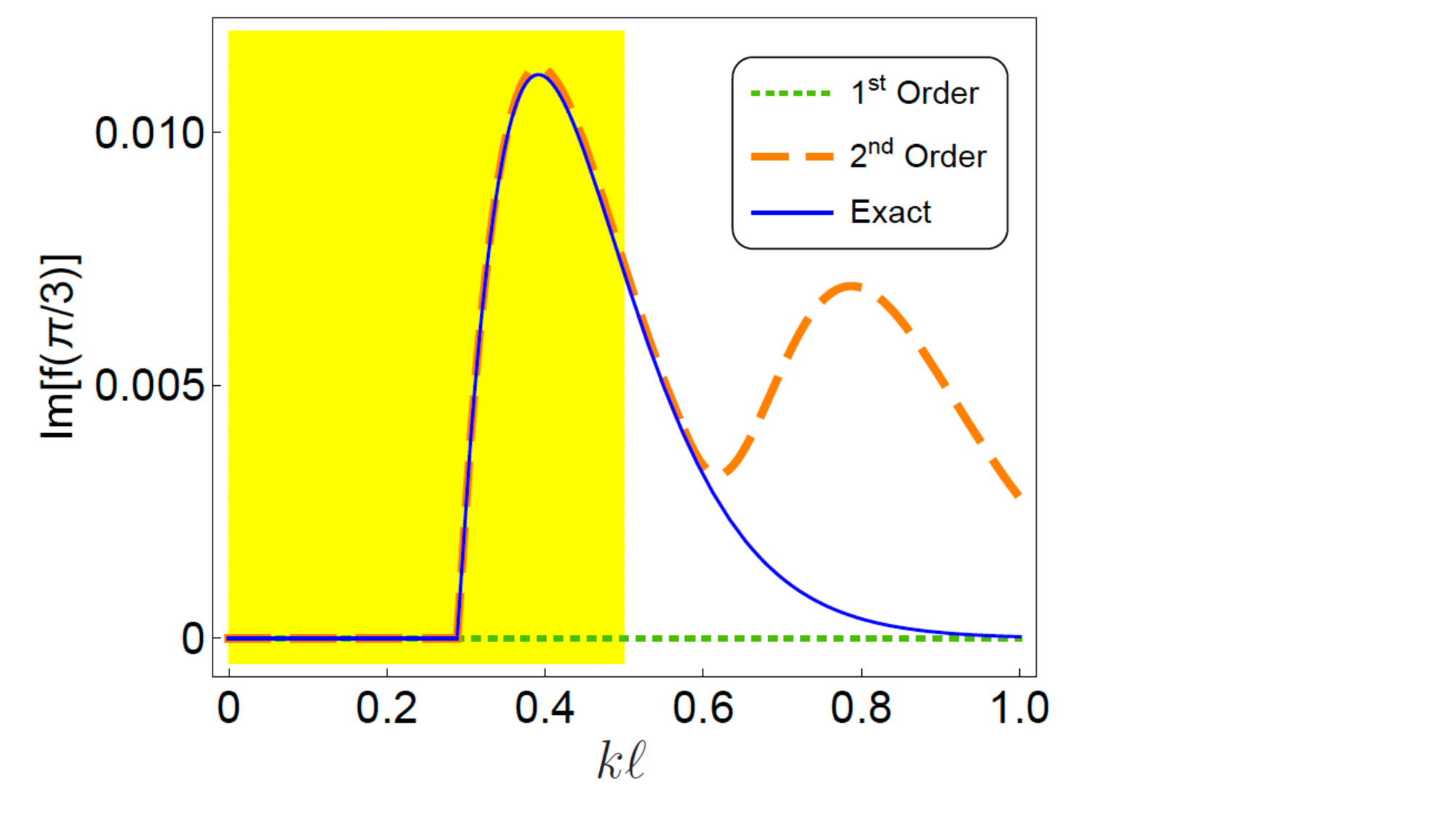}
	\caption{{Plots of the real and imaginary parts of the scattering amplitude $\ff(\pi/3)$ as a function of $k\ell$ for $\hat\varepsilon$ given by (\ref{epsilon=}) and (\ref{ex1}) with $\theta_0=4\pi/3$, $\fz=0.1$, $\alpha=500/{\rm mm}$, $\ell=1~\mu{\rm m}$, and $L=10~\mu{\rm m}$. The dotted and dashed curves respectively correspond to the first and second-order low-frequency approximations. The solid curve represents the exact expression which is valid for $k\ell\leq\alpha\ell=0.5$, i.e., in the region colored in yellow.}}
        \label{fig3}
        \end{center}
        \end{figure}

{Figure~\ref{fig3} shows the plots of the real and imaginary parts of $\ff(\theta)$ as a function of $k\ell$ for $\theta=\pi/3$, $\theta_0=4\pi/3$, $\fz=0.1$, $\alpha=500/{\rm mm}$, $\ell=1~\mu{\rm m}$, and $L=10~\mu{\rm m}$. Both the first- and second-order low-frequency approximations successfully describe the behavior of the real part of $\ff(\pi/3)$, while the former fails to provide a valid result for its imaginary part.\footnote{This is to be expected, because unlike $f(\theta)$, $f^{(1)}(\theta)$ takes real values.} For $k\leq\alpha$ where the exact result is valid, the discrepancy between the second-order approximation and the exact result is too small to be visible in these plots.}%

\section{{Low-frequency cloaking}}
\label{S5}

{According to Eqs.~\eqref{f=series-2D} -- \eqref{ff2=}, at low frequencies the slab is effectively invisible, i.e., it does not scatter the wave, if $\overline{\tilde w}_0(p;k)=\overline{\tilde w}_1(p,k)=0$, which in view of \eqref{w-ell=} means
	\be
	\int_0^\ell dx\,[\hat\varepsilon(x,y;k)-1]=\int_0^\ell dx\,x[\hat\varepsilon(x,y;k)-1]=0.
	\label{invisible-1}
	\ee
This observation suggests a simple cloaking procedure by coating the slab by a thin layer (or multilayer) of a material such that (\ref{invisible-1}) holds for the coated slab.}

{Suppose that the slab is coated by a thin bilayer with the following properties. The layers are made of homogeneous material with relative permittivities, $\hat\varepsilon_1$ and $\hat\varepsilon_2$, and thicknesses $\ell_1$ and $\ell_2$ which are bounded functions of $y$ and $k$. Then the relative permittivity of the coated slab has the form,
	\be
	\hat\varepsilon_c(x,y;k)=\left\{
	\begin{array}{ccc}
	\hat\varepsilon(x,y;k)&\for&0\leq x\leq\ell,\\[3pt]
	\hat\varepsilon_1(k) &\for&\ell<x\leq\ell+\ell_1(y;k),\\[3pt]
	\hat\varepsilon_2(k) &\for&\ell+\ell_1(y;k)<x\leq\ell+\ell_1(y;k)+\ell_2(y;k),\\[3pt]
	1 &&\rm{otherwise}.\end{array}\right.
	\label{coated-slab}
	\ee
Let $\ell_c\in\R^+$ be such that, $\ell+\ell_1(y;k)+\ell_2(y;k)\leq\ell_c\ll k^{-1}$ for all $y\in\R$. Then $k\ell_c\ll 1$ and our results apply to the coated slab. In particular, at low frequencies, it becomes effectively invisible provided that $\ell_1(y;k)$ and $\ell_2(y;k)$ are chosen in such a way that   (\ref{invisible-1}) holds with $\hat\varepsilon$ and $\ell$ changed to $\hat\varepsilon_c$ and $\ell_c$. In view of \eqref{coated-slab}, this condition is equivalent to the following pair of equations for $\ell_1$ and $\ell_2$.
	\begin{align}
	&\fz_1(k)\:\ell_1 +\fz_2(k)\:\ell_2 =-\ell\,\overline{w}_0(y;k),
	\label{eq1-5}\\
	&\fz_1(k)\:\ell_1 (\ell_1+2\ell)+
	\fz_2(k)\:\ell_2(\ell_2+2\ell_1+2\ell)=-2\ell^2\,\overline{w}_1(y;k),
	\label{eq2-5}
	\end{align}
where $\fz_{1,2}(k):=\hat\varepsilon_{1,2}(k)-1$, 
	\be
	\overline{w}_l(y;k):=\int_0^1 d\check x\: {\check x}^l\,w(\hat x,y;k)=
	\frac{1}{\ell^{l+1}}\int_0^\ell dx\: x^l[\hat\varepsilon(x,y;k)-1],
	\label{w-average}
	\ee
$l\in\{0,1\}$, and we have suppressed the dependence of $\ell_{1,2}$ on $y$ and $k$ for brevity.}

{Solving \eqref{eq1-5} for $\ell_1$ gives
	\be
	\ell_1(y;k)=-\frac{\fz_2(k)\:\ell_2(y;k)+\ell\,\overline{w}_0(y;k)}{\fz_1(k)}.
	\label{sol-1}
	\ee
Substituting this equation in \eqref{eq2-5} yields an equation for $\ell_2^2$ whose only real and positive solution is
	\be
	\ell_2(y;k)=\ell\,\sqrt{
	\frac{\overline{w}_0(y;k)^2+
	2\fz_1(k)[\overline{w}_1(y;k)-\overline{w}_0(y;k)]
	}{\fz_2(k)[\fz_2(k)-\fz_1(k)]}}
	\label{sol-2}
	\ee
provided that we can choose $\fz_1(k)$ and $\fz_1(k)$ such that the term in the square root takes a real and positive value. Otherwise, \eqref{eq1-5}  and  \eqref{eq2-5} have no real and positive solutions, and our method is not applicable for cloaking the slab. The same holds, if the right-hand side of \eqref{sol-1} fails to take a real and positive value.}

{Next, consider situations where the relative permittivity of the slab has the form,
	\be
	\hat\varepsilon(x,y;k)=\left\{\begin{array}{cc}
	1+\fz_0(k)g(y)  &\for~0\leq x\leq \ell,\\[3pt]
	1& {\rm otherwise},
	\end{array}\right.
	\label{finite-slab}
	\ee
where $\fz_0:\R^+\to\R$ and $g:\R\to\R$ are functions taking nonnegative values. Then, in view of  \eqref{w-average}, (\ref{sol-2}), and (\ref{finite-slab}), we have 
	\begin{align}
	&\overline w_0(y;k)=2\overline w_1(y;k)=\fz_0(k)\,g(y),
	\label{w-average-slab}\\
	&\ell_1(y;k)=-\frac{\ell[\fz_2(k)\sqrt{\cX(k,y)}+\fz_0(k)g(y)]}{\fz_1(k)},
	\label{L1=eg1}\\
	&\ell_2(y;k)=\ell\,\sqrt{\cX(k,y)},
	\label{L2=eg1}
	\end{align}
where
	\be 
	\cX(k,y):=\frac{\fz_0(k)g(y)[\fz_0(k)g(y)-\fz_1(k)]}{
	\fz_2(k)[\fz_2(k)-\fz_1(k)]}.
	\label{chi-X=}
	\ee
According to (\ref{L1=eg1}) -- (\ref{chi-X=}), $\ell_1$ and $\ell_2$ take real and positive values if $\fz_1(k)<0<\fz_2(k)$. This condition means that the first layer of coating should be made of a (meta)material with relative permittivity smaller than 1. This can be easily achieved using ordinary material if the slab is placed in a homogeneous background medium rather than in vacuum in which case $\varepsilon_0$ stands for {the background's} permittivity, and $\hat\varepsilon$ and $\hat\varepsilon_{1,2}$ are the relative permittivities relative to the background.}

\begin{figure}[!htbp]
        \begin{minipage}[c]{0.5\textwidth}
    	\caption{{Schematic view of a slab with a Gaussian permittivity modulation depicted in shades of blue and the corresponding low-frequency bilayer invisibility cloak. The pink and green regions represent the first and second layers with permittivities $1-\fz_0(k)$ and $1+0.4\fz_0(k)$, i.e., $\fz_1(k)=-\fz_0(k)$ and $\fz_2(k)=0.4\fz_0(k)$, respectively. The relative permittivity of the slab is given by \eqref{finite-slab} and (\ref{g=eq1}) with $L=2\ell$. The distances are measured in units of $\ell$.}} 
    	\label{fig4}
  	\end{minipage}\hfill
  	\hspace{2cm}
	\begin{minipage}[c]{0.4\textwidth}
    	\includegraphics[scale=.45]{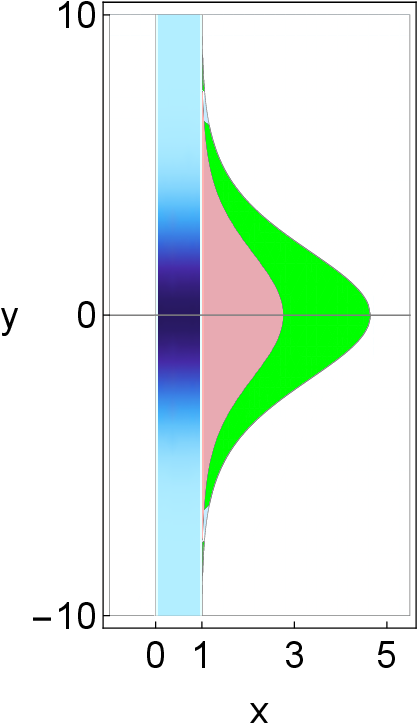}
  	\end{minipage}
  	\end{figure}

{Figure~\ref{fig4} shows the slab and the shape of its low-frequency bilayer invisibility cloak determined by \eqref{L1=eg1} and \eqref{L2=eg1} for 
	\be
	g(y)=e^{-y^2/2L^2}.
	\label{g=eq1}
	\ee
}        

\section{Dynamical formulation of low-frequency scattering in 3D}
\label{S6}

\subsection{DFSS in 3D}

\begin{figure}[ht]
	\begin{center}
        \includegraphics[scale=.30]{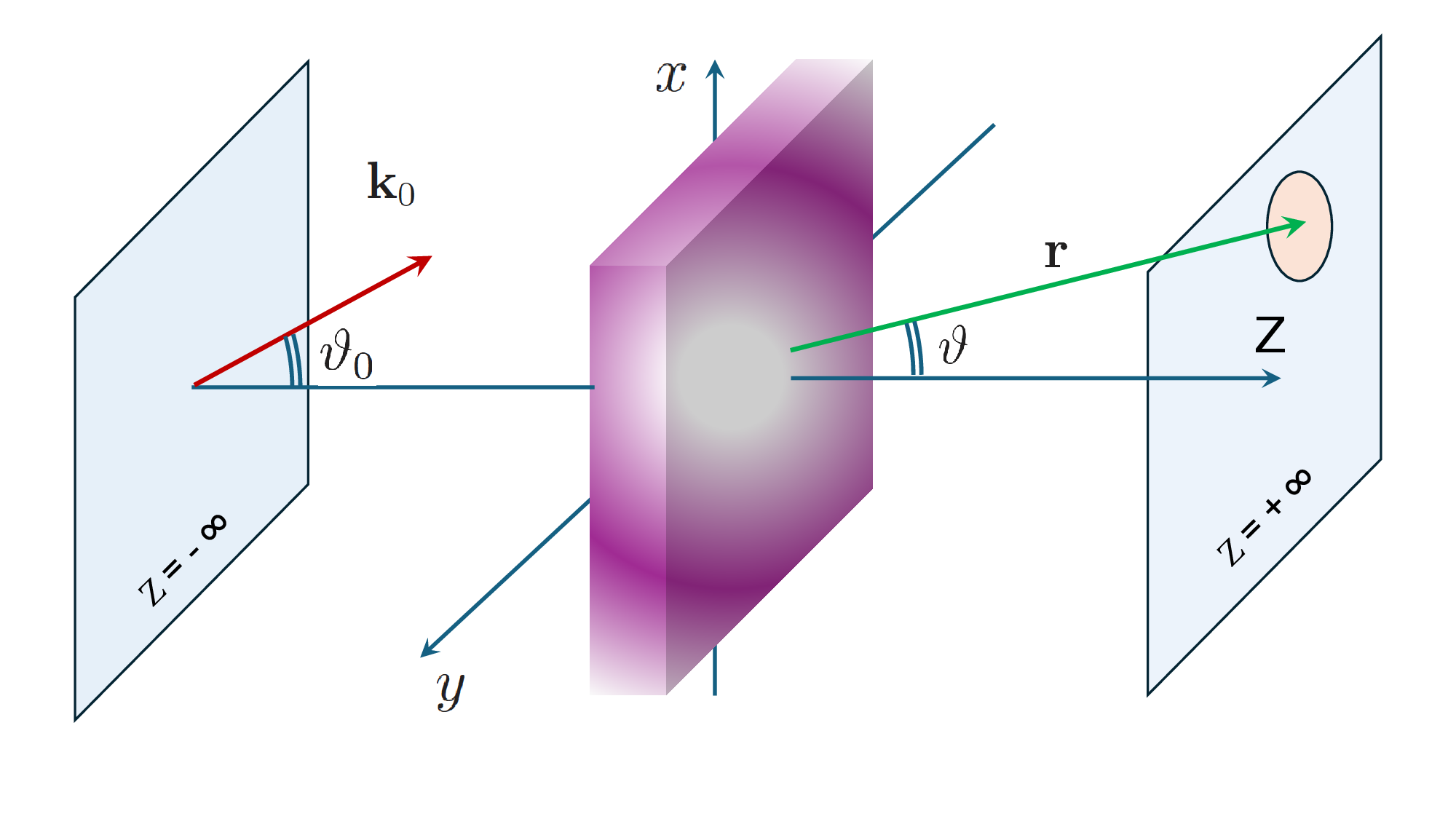}\vspace{-18pt}
	\caption{{Schematic view of the scattering setup with the source of the incidet wave located at $z=-\infty$. The blue planes represent the distant planes where the detectors are located. $\bfr$ marks the position of a detector screen that is placed at $z=+\infty$ and depicted as an orange ellipse.}} 
        \label{fig5}
        \end{center}
        \end{figure}

Following standard practice in 3D, we choose a coordinate system in which our slab is orthogonal to the $z$ axis, i.e., set $u=z$ so that the slab lies between the planes $z=0$ and $z=\ell${, as shown in Fig.~\ref{fig5}}. Furthermore, for each space vector $\bv\in\R^3$ we use an arrow to mark its projection onto the $x$-$y$ plane, i.e., set $\vec v:=(v_x,v_y)$ whenever $\bv=:(v_x,v_y,v_z)$. This allows us to employ the hybrid notation, $(\vec v,v_z):=(v_x,v_y,v_z)=\bv$, which simplifies some of the formulas. For example, for the position vector $\bfr=(x,y,z)$, we have $\vec r:=(x,y)$ and $\bfr=(\vec r,z)$.

Next, we identify $\sF^m$ and $\sF^m_k$ respectively with the space of functions (tempered distributions) $\bF:\R^2\to\C^{m\times 1}$ and the subspace of $\sF^m$ consisting of functions whose {supports lie} in the disk: 
	\[\sD_k:=\{\vec p\in\R^2~|~|\vec p|<k~\}.\] 
This means that  $\bF(\vec p)=0$ for all $\bF\in\sF^m_k$ and $|\vec p|\geq k$. We also let $\widehat{\vec r},\widehat{\vec p}:\sF^m\to\sF^m\times\sF^m$ and
$\widehat\varpi,\widehat\Pi_k:\sF^m\to\sF^m$ be the linear operators given by
	\begin{align}
	&(\widehat{\vec r}\;\bF)(\vec p):=
	i\vec{\nabla}_p\;F(\vec p):=i(\partial_{p_x},\partial_{p_y})\bF(\vec p),
	\quad\quad\quad\quad(\widehat{\vec p}\:\bF)(\vec p)=\vec p\,\bF(\vec p),\\[6pt]
	&\widehat\varpi:=\varpi(\widehat{\vec p}\,),
	\quad\quad\quad\quad\quad\varpi(\vec p\,):=\left\{\begin{array}{ccc}
	\sqrt{k^2-{\vec p\,}^2}&\for&|\vec p|<k,\\
	i\sqrt{{\vec p\,}^2-k^2}&\for&|\vec p|\geq k,\end{array}\right.
	\label{varpi-def}
	\\[6pt]
	&(\widehat\Pi_k \bF)(\vec p):=\left\{\begin{array}{ccc}
	\bF(\vec p)&\for&|\vec p|<k,\\
	\bzero&\for&|\vec p|\geq k,\end{array}\right.
	\end{align}
and $\vec p=(p_x,p_y)\in\R^2$. Note that in Dirac's bra-ket notation $\widehat\Pi_k$ takes the form,
	$\widehat\Pi_k=\int_{\sD_k} d^2\vec p\:|\vec p\kt\,\br\vec p|$, where $d^2\vec p:=dp_xdp_y$.
	
In Ref.~\cite{pra-2021}, we generalize the notion of fundamental transfer matrix to 3D and show that it admits a Dyson series expansion. For our slab system, it satisfies
	\be
	\widehat\bM=\widehat\Pi_k\sT e^{-i\int_0^\ell dz\:\widehat\bsH(z)}\:\widehat\Pi_k,
	\label{M=3D}
	\ee
where $\widehat\bsH(z)$ is given by (\ref{H=2D}) with $x$ changed to $z$, and $\widehat\varpi$ and $\widehat\sV(z)$ respectively given by (\ref{varpi-def}) and
	\begin{align}
	\widehat\sV(z):=v(\widehat{\vec r},z;k),
	&&v(\vec r,z;k):=k^2[1-\hat\varepsilon(\vec r,z;k)].
	\end{align}
Again it is easy to establish the following 3D analog of (\ref{V=}).
	\be
	\left(\widehat\sV(z)\bF\right)(\vec p)=\frac{1}{4\pi^2}\int_{\R^2} d^2\vec q~\tilde{\tilde v}(\vec p-\vec q,z;k)\bF(\vec q),
	\label{V=3D}
	\ee
where $\tilde{\tilde v}(\vec p,z;k)$ stands for the 2D Fourier transform of 
$v(\vec r,z;k)$ with respect to $\vec r$, i.e., 
	\[\tilde{\tilde v}(\vec p,z;k):=\int_{\R^2}d^2\vec r\:e^{-i\vec r\cdot\vec p}v(\vec r,z;k).\]
	
Next, we recall the definition of the scattering amplitude $\ff$ in 3D. We use conventions where it is given through the following asymptotic expression for the scattering solutions {\cite{yafaev}} of the Helmholtz equation~(\ref{H-eq}) in 3D.
	\be
	\psi(\bfr)\to\frac{1}{(2\pi)^{3/2}}\left[e^{i\bk_0\cdot\bfr}+\frac{e^{ikr}}{r}\,\ff(\vartheta,\varphi)\right]~~~\for~~~z\to\pm\infty,
	\ee
where $r,\vartheta$, and $\varphi$ are respectively the spherical radial, polar, and azimuthal coordinates of $\bfr$. 

It is useful to introduce the scattered wave vector, $\bk:=k\, r^{-1}\bfr$, which has spherical coordinates $(k,\vartheta,\varphi)$. Similarly, we denote the spherical coordinates of the incident wave vector $\bk_0$ by $(k,\vartheta_0,\varphi_0)$. Then in analogy with 2D, we can speak of left-incident  (respectively right-incident) waves whose source is located at $z=-\infty$ (respectively $z=+\infty$). In particular, we can identify the left- and right-incident waves through the conditions, $\cos\vartheta_0>0$ and $\cos\vartheta_0<0$, respectively. We also introduce the scattering amplitudes for the left- and right-incident waves $\ff^{l/r}$ which satisfy
	\be
	\ff(\vartheta,\varphi)=\left\{\begin{array}{ccc}
	\ff^l(\vartheta,\varphi)&\for&\cos\vartheta_0>0,\\
	\ff^r(\vartheta,\varphi)&\for&\cos\vartheta_0<0,
	\end{array}\right.
	\label{ff=3D}
	\ee
as well as the following 3D analogs of (\ref{f-Left}) and (\ref{f-Right}), \cite{pra-2021}.
	\begin{align}
	\ff^l(\vartheta,\varphi)=\frac{-i}{\sqrt{2\pi}}\times\left\{
	\begin{array}{ccc}
	A^l_+(\vec k)-\check\delta(\vec k-\vec k_0)&\for&
	\cos\vartheta>0,\\
	B^l_-(\vec k)&\for&
	\cos\vartheta<0,\end{array}\right.
	\label{f-Left=3}\\[6pt]
	\ff^r(\vartheta,\varphi)=\frac{-i}{\sqrt{2\pi}}\times\left\{
	\begin{array}{ccc}
	A^r_+(\vec k)&\for&
	\cos\vartheta>0,\\
	B^r_-(\vec k)-\check\delta(\vec k-\vec k_0)&\for&
	\cos\vartheta<0,
	\end{array}\right.
	\label{f-Right=3}
	\end{align}
where $A^{l/r}_+,B^{l/r}_-\in\sF_k^1$ fulfil
	\begin{align}
	&A^l_+=\widehat M_{12}B^l_-+\widehat M_{11}\check\delta_{\vec k_0},
	\label{A-L=3}\\
	& \widehat M_{22} B^l_-=-\widehat M_{21}\check\delta_{\vec k_0},
	\label{B-L=3}\\
	&A^r_+=\widehat M_{12}B^r_-,
	\label{A-R-=3}\\
	&\widehat M_{22} B^r_-=\check\delta_{\vec k_0},
	\label{B-R=3}
	\end{align}
$\vec k$ and $\vec k_0$ are respectively the projections of the scattered and incident wave vectors onto the $x$-$y$ plane, which are given by $\vec k:=k\sin\vartheta(\cos\varphi,\sin\varphi)$ and $\vec k_0:=k\sin\vartheta_0(\cos\varphi_0,\sin\varphi_0)$, 
and 
	\begin{align}
	&\check\delta(\vec p):=4\pi^2\varpi(\vec k_0)\delta(\vec p)
	=4\pi^2 k |\cos\vartheta_0|\delta(\vec p),	
	\nn
	&\check\delta_{\vec k_0}(\vec p):=\check\delta(\vec p-\vec k_0),
	\quad\quad\quad
	\delta(\vec p):=\delta(p_x)\delta(p_y).\nn
	\end{align} 

Because (\ref{A-L=3}) -- (\ref{B-R=3}) have the same structure as (\ref{A-L}) -- (\ref{B-R}), we can again introduce the linear operators $\widehat N_{ab}:\sF^1_k\to\sF^1_k$ using (\ref{N-ab=}) and obtain formal series solutions for (\ref{A-L=3}) -- (\ref{B-R=3}). This leads to (\ref{B-L-series}) -- (\ref{A-R-series}) provided that we make the following changes in these equations.
	\begin{align}
	&p\to\vec p,
	&&\varpi(p_0)\to 2\pi\varpi(\vec k_0),
	&&p_0\to\vec k_0.
	\label{trans-3D}
	\end{align}

\subsection{Low-frequency scattering in 3D}

Suppose that 
	\be
	\hat\varepsilon(\vec r,z;k)=1+\left\{\begin{array}{ccc}
	w(\vec r,\frac{z}{\ell};k) &\for& x\in[0,\ell],\\
	0&\for&x\notin[0,\ell],
	\end{array}\right.
	\label{epsilon=3D}
	\ee
for some function $w:\R^2\times[0,\ell]\times\R^+\to\C$. Then,
	\begin{align}
	&v(\vec x,z;k)=\left\{\begin{array}{ccc}
	-k^2w(\vec r,\frac{z}{\ell},y;k) &\for& x\in[0,\ell],\\
	0&\for&x\notin[0,\ell],
	\end{array}\right.
	&&\widehat\sV(x)=-k^2\widehat\sW(\mbox{\large$\frac{z}{\ell}$}),
	\label{cV=3D}
	\end{align}
where 
	\begin{align}
	&\widehat\sW(\check z):=\left\{\begin{array}{ccc}
	w(\widehat{\vec r},\check z;k)&\for& \check z\in[0,1],\\[3pt]
	\widehat 0&\for&\check z\notin[0,1].
	\end{array}\right.
	\label{cW=3D}
	\end{align}
Comparing (\ref{cV=3D}) and (\ref{cW=3D}) with their 2D analogs, namely (\ref{cV=}) and (\ref{cW=}), and making use of the structural similarity between the Dyson series expansions of the transfer matrices in 2D and 3D, we can repeat the analysis of Sec.~\ref{S3} to show that $\widehat M_{ab}$ satisfies (\ref{Mab=3}) with $\widehat{{\check\cR}}$ given by (\ref{Q=2}). This in turn implies that (\ref{Nab=3}) holds also in 3D. The rest of the analysis of Sec.~\ref{S3} applies as well, and we find the following 3D analogs of (\ref{z11}) --(\ref{W-2-series}) and (\ref{W-j-n=}) -- (\ref{w-ell=}). 
		\begin{align}
		&e^{(-1)^a i\check x_2\ell\widehat\varpi}\widehat\Pi_k=
		\int_{\sD_k}d^2\vec p\:e^{(-1)^a i\check x_2 k\ell\sqrt{1-\vec p^{\,2}/k^2}}\,
		|\vec p\kt\,\br \vec p|,
		\label{z11-3D}\\
		&\widehat\Pi_k\,\widehat\varpi^{-1}e^{(-1)^{b-1}i\check x_1\ell\widehat\varpi}=
		\frac{1}{k}
		\int_{\sD_k}d^2{\vec p\,}'\: \frac{e^{(-1)^{b-1}i\check x_1 k\ell\sqrt{1-{{\vec p\,}'}^{2}/k^2}}}{\sqrt{1-{\vec p\,}^{'2}/k^2}}\,|{\vec p\,}'\kt\,\br {\vec p\,}'|,\\
		&\br \vec p|\widehat\sW(\check x)|{\vec p\,}'\kt=\frac{\tilde{\tilde w}(\vec p-{\vec p\,}',\check x;k)}{4\pi^2},\\[6pt]
		&\widehat\sW_2(\check x,\check x')=\sum_{j=0}^\infty s_j(\check x-\check x')
		\widehat\sW(\check x)[(k\ell)^2-\ell^2{\widehat{\vec p}}^{\,2}]^j\widehat\sW(\check x'),
		\label{W-2-series=3D}
		\end{align}
		\begin{align}
		&\widehat\sW_{j_1,j_2,\cdots,j_{n-1}}(\check x_n,\check x_{n-1},\cdots,\check x_1):=
		\prod_{r=1}^{n-1}
		\widehat\sW(\check x_{n-r+1})[(k\ell)^2-\ell^2\widehat{\vec p}^{\,2}]^{j_{n-r}}
		\,\widehat\sW(\check x_1).
		\label{W-j-n=3D}\\
	&\widehat N_{ab}=\sum_{j=1}^\infty (k\ell)^{j}
	\int_{\sD_k}d^2\vec p
	\int_{\sD_k}d^2{\vec p\,}'\: N^{(j)}_{ab}(\vec p,{\vec p\,}';k)|\vec p\kt\,\br {\vec p\,}'|,
	\label{N-ab=2-3D}\\
	&N^{(1)}_{ab}(\vec p,{\vec p\,}';k)=\frac{(-1)^a\, i~ \overline{\tilde{\tilde w}}_0(\vec p-{\vec p\,}';k)}{8\pi^2\sqrt{1-{\vec p\,}^{'2}/k^2}},
	\label{N1=3D}\\
	&N^{(2)}_{ab}(\vec p,{\vec p\,}';k)=\frac{1}{8\pi^2}
	\left[(-1)^{a+b}-\sqrt{\frac{k^2-{\vec p\,}^2}{k^2-{\vec p\,}^{'2}}}\right]
	\:\overline{\tilde{\tilde w}}_1(\vec p-{\vec p\,}';k),
	\label{N2=3D}\\
	&\overline{\tilde{\tilde w}}_l(\vec p;k):=\int_0^1d\check z\: \check z^l\:\tilde{\tilde w}(\vec p,\check z;k).
	\label{w-ell=3D}
	\end{align}

Performing the transformation (\ref{trans-3D}) in (\ref{B-L-series}) -- (\ref{A-R-series}) and using the resulting equations together with (\ref{Q=2}), (\ref{Nab=3}), (\ref{ff=3D}) -- (\ref{f-Right=3}) and (\ref{z11-3D}) -- (\ref{N2=3D}), we find
	\be
	\ff(\vartheta,\varphi)=	\ff^{(1)}(\vartheta,\varphi)\,k\ell+\ff^{(2)}(\vartheta,\varphi)\,(k\ell)^2+\cO(k\ell)^3,
	\label{f-=expand-3D}
	\ee
where
	\begin{align}
	&\ff^{(1)}(\vartheta,\varphi):=
	\frac{k}{2\sqrt{2\pi}}\:
	\overline{\tilde{\tilde w}}_0\big(k\vec \fg(\vartheta,\varphi,\vartheta_0,\varphi_0);k\big),
	\label{ff1=3D}\\
	&\begin{aligned}
	\ff^{(2)}(\vartheta,\varphi)&:=
	\frac{ik}{2\sqrt{2\pi}}\Big[
	(\cos\vartheta_0-\cos\vartheta)\,
	\overline{\tilde{\tilde w}}_1\big(k\vec \fg(\vartheta,\varphi,\vartheta_0,\varphi_0);k\big)+
	\\&\hspace{1.9cm}
	\frac{k^2}{8\pi^2}\int_0^{\frac{\pi}{2}}\!\!d\alpha\int_0^{2\pi}\!\!d\beta\:\sin\alpha\:
	\overline{\tilde{\tilde w}}_0\big(k\vec \fg(\vartheta,\varphi,\alpha,\beta);k\big)\,
	\overline{\tilde{\tilde w}}_0
	\big(k\vec \fg(\alpha,\beta,\vartheta_0,\varphi_0);k\big)\Big],
	\end{aligned}
	\label{ff2=3D}\\
	&\vec \fg(\vartheta,\varphi,\alpha,\beta):=(\sin\vartheta\cos\varphi-\sin\alpha\cos\beta\,,\,
	\sin\vartheta\sin\varphi-\sin\alpha\sin\beta).
	\end{align}
	
As a particular example, consider taking $w(\vec r,\check z;k)=\fz\,e^{-{\vec r\,}^2/2L^2}$, where $\fz$ and $L$ are respectively complex and real parameters, and $L>0$. Then, 
	\be
	\hat\varepsilon(x,y,z;k)=
	\left\{\begin{array}{ccc}
	1+\fz\,e^{-(x^2+y^2)/2L^2}&\for&z\in[0,\ell],\\
	1&\for&z\notin[0,\ell],\end{array}\right.
	\label{3D-gaussian1}
	\ee
and \eqref{w-ell=3D}, \eqref{ff1=3D}, and \eqref{ff2=3D} give
	\begin{align}
	&\overline{\tilde{\tilde w}}_0(\vec p;k)=2\overline{\tilde{\tilde w}}_1(\vec p;k)=2\pi\fz\,L^2
	e^{-\frac{1}{2}L^2{\vec p\,}^2},\\
	&\ff^{(1)}(\vartheta,\varphi)=\sqrt{\frac{\pi}{2}}\frac{\fz\,\fK^2}{k}\,
	e^{\fK^2 h(\vartheta,\varphi-\varphi_0,\vartheta_0)},\\
	&\ff^{(2)}(\vartheta,\varphi)=\sqrt{\frac{\pi}{2}}\frac{i\fz\,\fK^2}{2k}
	\Big[(\cos\vartheta_0-\cos\vartheta)\,e^{\fK^2 h(\vartheta,\varphi-\varphi_0,\vartheta_0)}+\fz\,\fK^2\cY(\vartheta,\varphi,\vartheta_0,\varphi_0,\fK)\Big],
	\end{align}
where $\fK:=kL$ and 
	\begin{align}
	&h(\vartheta,\varphi,\vartheta_0):=\sin\vartheta_0\sin\vartheta\cos\varphi+\frac{1}{4}(
	\cos2\vartheta+\cos2\vartheta_0)-\frac{1}{2},\nn\\
	&\cY(\vartheta,\varphi,\vartheta_0,\varphi_0,\fK):=
	\frac{1}{2\pi}\int_0^{\frac{\pi}{2}}\!\!d\alpha\int_0^{2\pi}\!\!d\beta\: \sin\alpha\;e^{\fK^2[h(\vartheta,\varphi-\beta,\alpha)+h(\alpha,\beta-\varphi_0,\vartheta_0)]}.
	\end{align}	
	
\begin{figure}
	\begin{center}
        	\includegraphics[scale=.3]{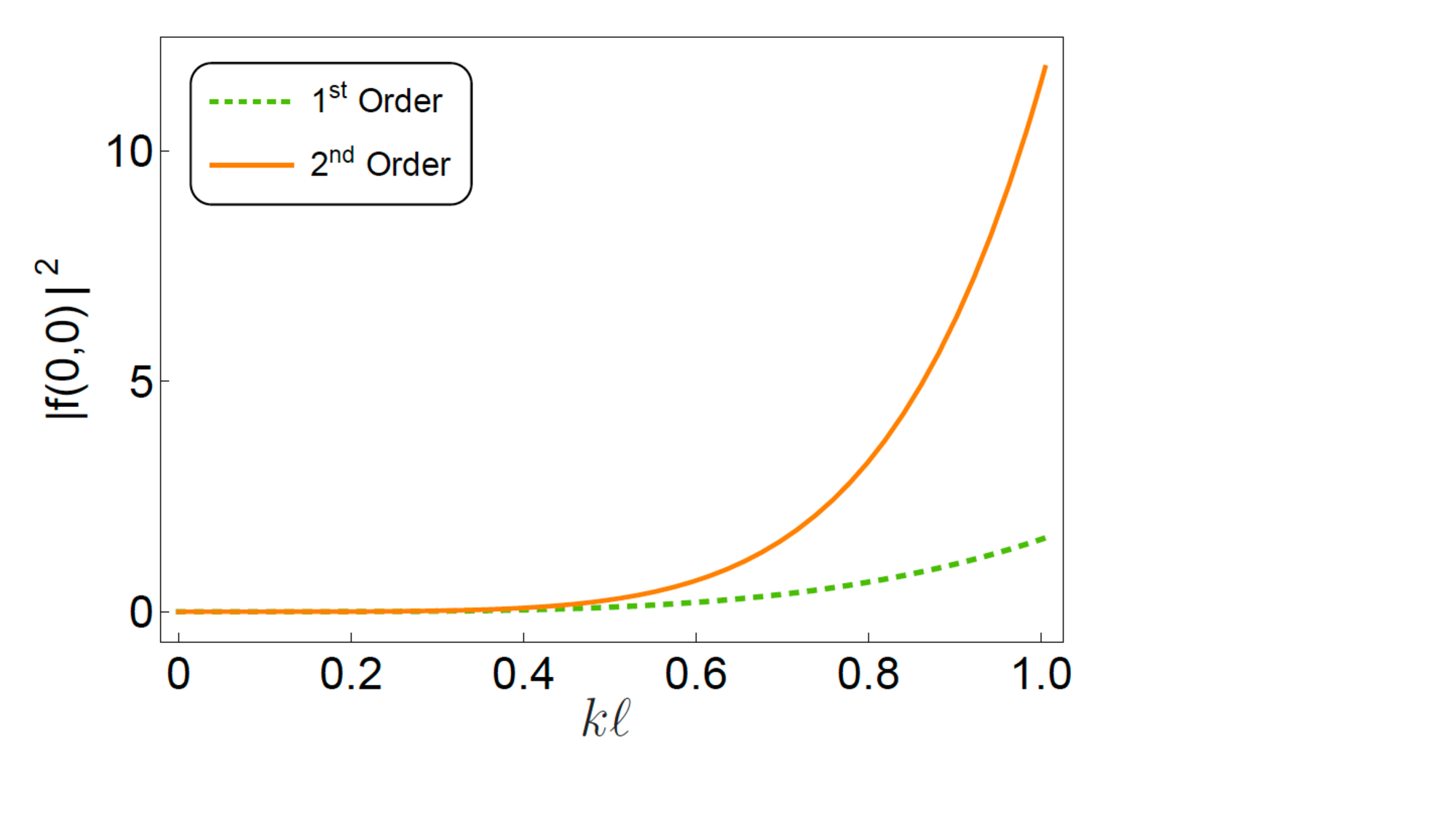}~\hspace{-1.7cm}~\includegraphics[scale=.3]{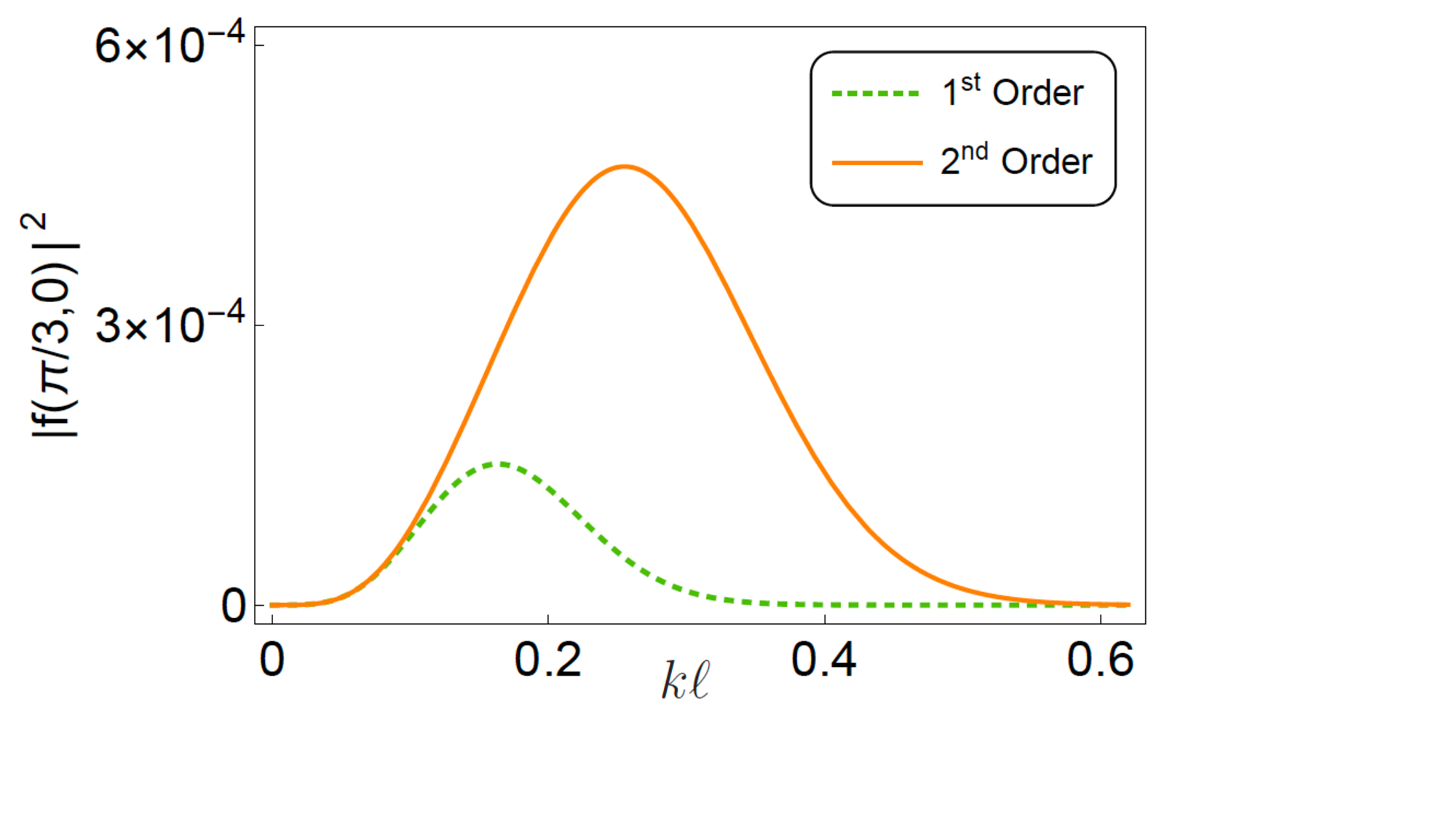}\\
         \includegraphics[scale=.3]{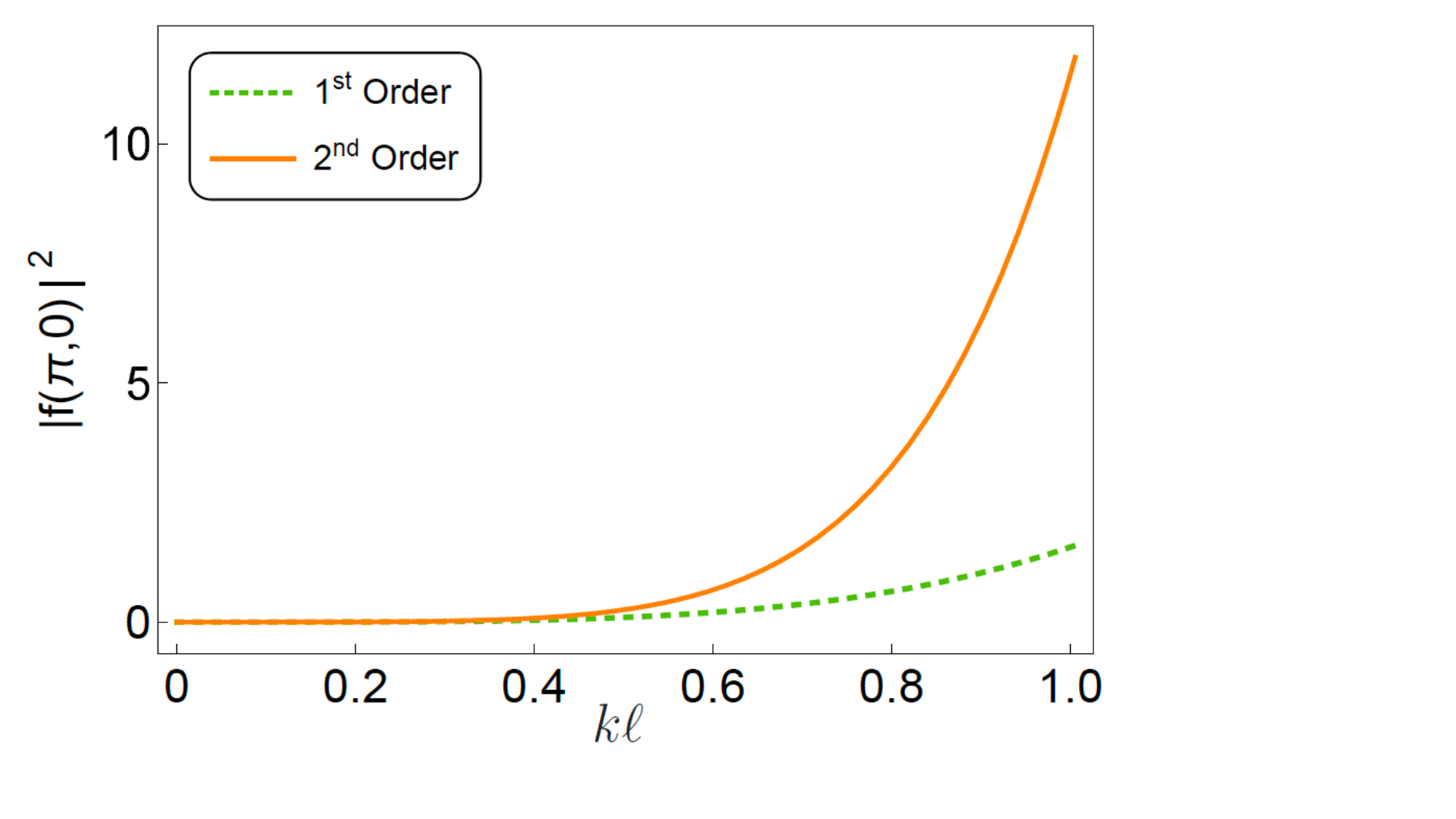}~\hspace{-1.7cm}~\includegraphics[scale=.3]{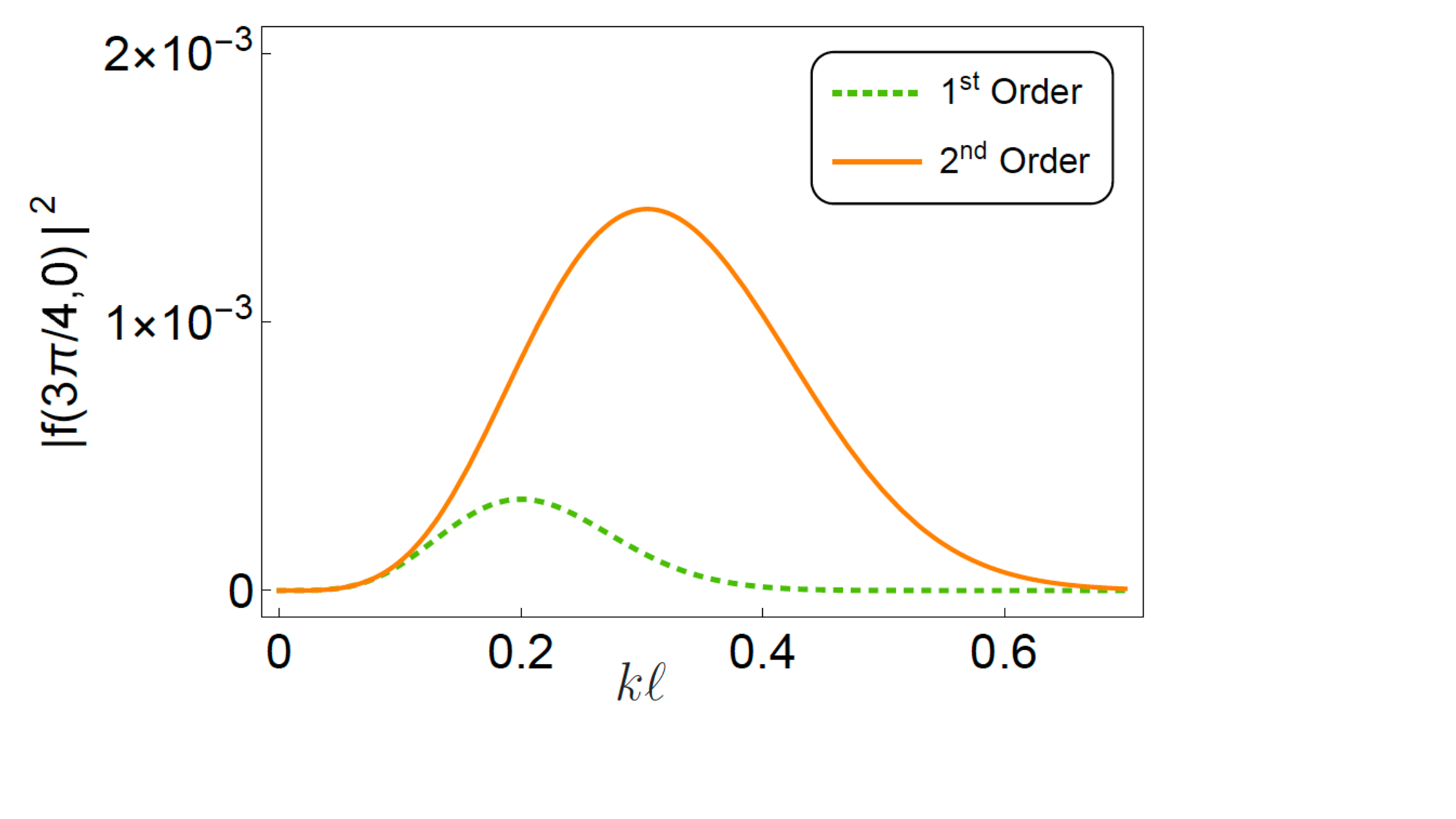}
	\caption{{Plots of the scattering cross section {$|\ff(\vartheta,0)|^2$ in units of ${\rm mm}^2$ as a function of $k\ell$ for the permittivity profile (\ref{3D-gaussian1}) obtained using the first and second-order low-frequency approximations for $\vartheta=0$ (top left), $\pi/3$ (top right), $3\pi/4$ (bottom right), $\pi$ (top left). The other parameters of the system are given by~(\ref{spec-gaussian}).}}}
        \label{fig6}
        \end{center}
        \end{figure}

Figure~\ref{fig6} shows the plots of the scattering cross section $|\ff(\vartheta,0)|^2$ as a function of the wave number $k$ for 
	\begin{align}
	&\vartheta_0=\varphi_0=0,
	&&\fz=10,
	&&\ell=1~{\rm mm},
	&&L=10~{\rm mm},
	\label{spec-gaussian}
	\end{align}
and different values of $\vartheta$.
{Plotting $|\ff(\vartheta,0)|^2$ as a function of $k$ for other values of $\vartheta$ produces curves that have similar features to those of $|\ff(\frac{\pi}{3},0)|^2$ and  $|\ff(\frac{3\pi}{4},0)|^2$. The discrepancy between the first- and second-order low-frequency approximations is more pronounced for values of $\vartheta$ that are closer to $\pi/2$.  For $\vartheta=0$ and $\pi$, i.e., forward and backward scattering, they produce identical results for low frequencies. }
        
\begin{figure}
	\begin{center}
        \includegraphics[scale=.5]{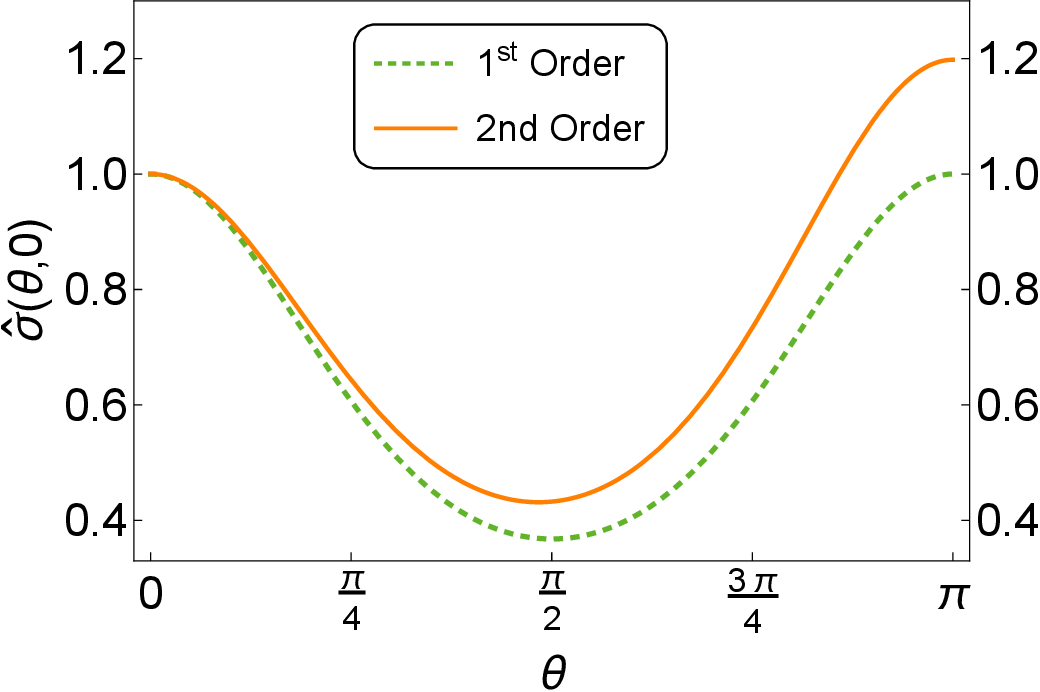} 
	\caption{{Plots of $\hat\sigma(\vartheta,0)$ for the permittivity profile given by (\ref{3D-gaussian1}), (\ref{spec1-3D}), and $kL=1$ that are obtained using the first- and second-order low-frequency approximations.}}
        \label{fig7}
        \end{center}
        \end{figure}

\begin{figure}
	\begin{center}
	\includegraphics[scale=.55]{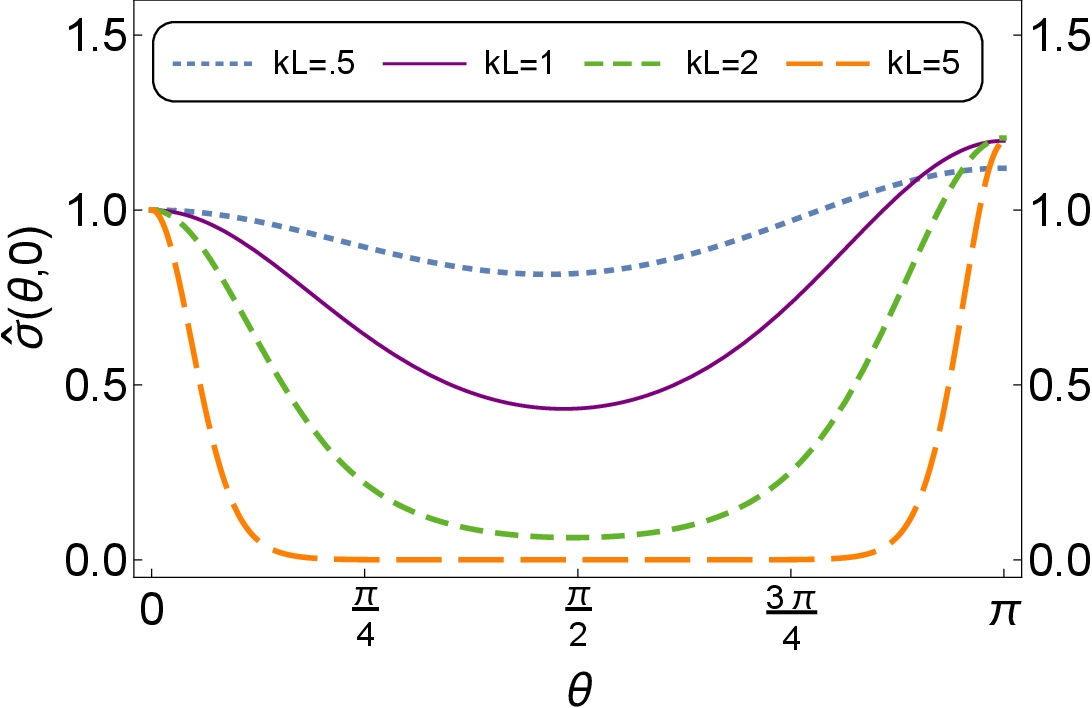} 
	\caption{Plots of $\hat\sigma(\vartheta,0)$ obtained using second-order low-frequency approximation for the permittivity profile given by (\ref{3D-gaussian1}) and (\ref{spec1-3D}) for different values of $kL$.}
        \label{fig8}
        \end{center}
        \end{figure}

{Figure~\ref{fig7} shows the plots of the normalized scattering cross section, 
	\be
	\hat\sigma(\vartheta,\varphi):=\frac{|\ff(\vartheta,\varphi)|^2}{|\ff(0,0)|^2},
	\label{normalized-3D}
	\ee 
as a function of $\vartheta$ that are obtained using the first- and second-order low-frequency approximations for the following values of the other parameters of the scattering problem.}
	\begin{align}
	&\varphi=\vartheta_0=\varphi_0=0,
	&&\fz=10,
	&&k\ell=0.2,
	\label{spec1-3D}
	\end{align}
and $kL=1$.
Figure~\ref{fig8} shows the plots of $\hat\sigma(\vartheta,0)$ as a function of $\vartheta$ for different values of $kL$.	
   {As seen from these graphs, the intensity of the scattered wave arriving at $z=-\infty$, where the source of the incident wave is located, is larger than that of the scattered wave reaching $z=+\infty$. For larger values of $L$ the scattering is dominated by the forward and backward scattering.}
   
   {Examining the analogs of Figs.~\ref{fig7} and \ref{fig8} for other small values of $k\ell$, we find curves with identical structures to those given in these figures. Therefore, changing the value $k\ell$ does not lead to any qualitative differences in the behavior of the scattering cross section.}

\section{Concluding remarks}
\label{S7}

Transfer matrices have been used in performing scattering calculations for several decades. The recent discovery that they can be expressed in terms of the time-evolution operators for certain non-unitary quantum systems \cite{ap-2014} revealed some of their previously unrecognized features in one dimension \cite{tjp-2020}. This observation paved the way towards developing a dynamical formulation of stationary scattering for scalar waves in two and three dimensions \cite{pra-2021} and electromagnetic waves \cite{pra-2023}. In this article we have explored the application of this formulation  in constructing the low-frequency series expansion of the scattering amplitude in two and three dimensions. This was mainly motivated by our earlier investigation of low-frequency scattering in one dimension  \cite{jmp-2021, jpa-2021}. 

The central ingredients of our approach to low-frequency scattering in two and three dimensions are the Dyson series for the fundamental transfer matrix $\widehat\bM$ and the series solutions of the integral equations that yield the scattering amplitude $\ff$ in terms of the entries of $\widehat\bM$. Making use of these ingredients, we have been able to outline a method for calculating the coefficients of the low-frequency series expansion of $\ff$. This turns out to be quite easy for the leading-order and next-to-leading-order terms and, similarly to one dimension, yield closed-form analytic expressions for these terms. {Neglecting the higher order terms gives rise to a low-frequency approximation. To assess the reliability of this approximation, we have examined its utility in the study of a class of exactly solvable scattering problems. We have verified the perfect agreement between our approximate results and the exact result at low {frequencies} by explicit analytic calculations.}

{The analytic expressions for the low-frequency scattering amplitude may be easily used for the purpose of designing inhomogeneous thin films with specific scattering features at low frequencies. For example they allow for identifying a specific coating of the scatterer that reduces its scattering effects appreciably at low frequencies. This yields a simple scheme for devising a low-frequency invisibility cloak. }

An important advantage of our method over those employed in the earlier studies of this problem is that we identify the term ``low-frequency wave'' with the requirement $k\ell\ll 1$, where $k$ is the wave number and $\ell$ is the characteristic length scale describing the spatial extension of the scatterer along the scattering axis. This means that our method applies also to situations where the wavelength of the incident wave is comparable or even smaller than the {spatial} extension of the scatterer in the normal directions to the scattering axis.

\section*{Acknowledgements}
This work has been supported by the Scientific and Technological Research Council of Turkey (T\"UB\.{I}TAK) in the framework of the project 123F180 and by Turkish Academy of Sciences (T\"UBA). We wish to express our gratitude to Emrah Y\"ukselci for his numerical confirmation of our approximate analytic results for a simple toy model.

\section*{{Appendix: Derivation of Eq.~(\ref{compact})}}

{We begin our derivation of Eq.~(\ref{compact}) by noting that in view of (\ref{K=}), $\bcK\, e^{ix\widehat\varpi\bsigma_3}\bcK=2i\sin(x\widehat\varpi)\,\bcK$. With the help of this identity and (\ref{H=2D}), we can show that, for $n\geq 2$ and $x_1\leq x_2\leq\cdots\leq x_n$,
	\begin{align}
	\widehat\bsH(x_n)\widehat\bsH(x_{n-1})\cdots\widehat\bsH(x_1)=&
	\frac{i^{n-1}}{2}\, e^{-ix_n\widehat\varpi\bsigma_3}\:
	\widehat\sV(x_n)\widehat\sV(x_n,x_{n-1},\cdots,x_1)\,\widehat\varpi^{-1}\bcK\:
	e^{ix_1\widehat\varpi\bsigma_3},
	\label{H-x's=}
	\end{align}
where
	\begin{align}
	\widehat\sV(x_n,x_{n-1},\cdots,x_1):=
	\widehat s(x_n-x_{n-1})\widehat\sV(x_{n-1})\widehat s(x_{n-1}-x_{n-2})
	\widehat\sV(x_{n-2})\cdots	\widehat s(x_2-x_1)\widehat\sV(x_1).
	\label{W=}
	\end{align}
Eqs.~(\ref{step}), (\ref{s-x=}), (\ref{H-x's=}), and (\ref{W=}) allow us to identify 
	\be
	\int_{x_0}^x dx_n\int_{x_0}^{x_n}dx_{n-1}\cdots\int_{x_0}^{x_2}dx_1
	\widehat\bsH(x_n)\widehat\bsH(x_{n-1})\cdots\widehat\bsH(x_1)
	\label{z1}
	\ee
with 
	\be
	\frac{i^{n-1}}{2}
	\int_{x_0}^x dx_n\int_{x_0}^xdx_{n-1}\cdots\int_{x_0}^x dx_1
	e^{-ix_n\widehat\varpi\bsigma_3}\:
	\widehat\sV(x_n)\widehat\sV(x_n,x_{n-1},\cdots,x_1)\,\widehat\varpi^{-1}\bcK\:
	e^{ix_1\widehat\varpi\bsigma_3}.
	\label{z2}
	\ee
Substituting (\ref{z2}) for (\ref{z1}) in (\ref{dyson-2D}) and making use of (\ref{step}), (\ref{s-x=}), (\ref{V-x=}), (\ref{S=}), (\ref{W=}), and the completeness relation, $\int_{-\infty}^\infty dx\:|x\kt\,\br x|=\widehat I$, we obtain \eqref{compact}.}

\ed